\documentclass[twocolumn,twocolappendix]{aastex63}

\usepackage{listings}
\usepackage{amsmath}
\usepackage{hyperref}

\graphicspath{{./}{figures/}}
\shorttitle{Assembly Bias from DESI Counts-in-Cylinders}

\shortauthors{Pearl, et al.}

\newcommand{\wprp}{$w_{\rm p}(r_{\rm p})$}

\begin{document}

\title{The DESI One-Percent Survey: Evidence for Assembly Bias from Low-Redshift Counts-in-Cylinders Measurements}

\author[0000-0001-9820-9619]{Alan N.\ Pearl}
\affiliation{Department of Physics and Astronomy, University of Pittsburgh, Pittsburgh, PA 15260, USA}
\affiliation{Pittsburgh Particle Physics, Astrophysics, and Cosmology Center (PITT PACC), University of Pittsburgh, Pittsburgh, PA 15260, USA}

\author[0000-0002-6443-7186]{Andrew R.\ Zentner}
\affiliation{Department of Physics and Astronomy, University of Pittsburgh, Pittsburgh, PA 15260, USA}
\affiliation{Pittsburgh Particle Physics, Astrophysics, and Cosmology Center (PITT PACC), University of Pittsburgh, Pittsburgh, PA 15260, USA}

\author[0000-0001-8684-2222]{Jeffrey A.\ Newman}
\affiliation{Department of Physics and Astronomy, University of Pittsburgh, Pittsburgh, PA 15260, USA}

\author[0000-0001-5063-8254]{Rachel Bezanson}
\affiliation{Department of Physics and Astronomy, University of Pittsburgh, Pittsburgh, PA 15260, USA}

\author[0000-0001-7690-2260]{Kuan Wang}
\affiliation{Department of Physics, University of Michigan, Ann Arbor, MI 48109, USA}
\affiliation{Leinweber Center for Theoretical Physics, University of Michigan, Ann Arbor, MI 48109, USA}

\author[0000-0002-2733-4559]{John Moustakas}
\affiliation{Department of Physics and Astronomy, Siena College, 515 Loudon Road, Loudonville, NY 12211, USA}

\author[0000-0003-0822-452X]{Jessica N.\ Aguilar}
\affiliation{Lawrence Berkeley National Laboratory, 1 Cyclotron Road, Berkeley, CA 94720, USA}

\author[0000-0001-6098-7247]{Steven Ahlen}
\affiliation{Boston University, 590 Commonwealth Avenue, Boston, MA 02215, USA}

\author{David Brooks}
\affiliation{Department of Physics \& Astronomy, University College London, Gower Street, London, WC1E 6BT, UK}

\author{Todd Claybaugh}
\affiliation{Lawrence Berkeley National Laboratory, 1 Cyclotron Road, Berkeley, CA 94720, USA}

\author[0000-0002-5954-7903]{Shaun Cole}
\affiliation{1Institute for Computational Cosmology, Department of Physics, Durham University, South Road, Durham DH1 3LE, UK}

\author{Kyle Dawson}
\affiliation{Department of Physics and Astronomy, The University of Utah, 115 South 1400 East, Salt Lake City, UT 84112, USA}

\author[0000-0002-1769-1640]{Axel de la Macorra}
\affiliation{Instituto de F\'{\i}sica, Universidad Nacional Aut\'{o}noma de M\'{e}xico,  Cd. de M\'{e}xico  C.P. 04510,  M\'{e}xico}

\author{Peter Doel}
\affiliation{Department of Physics \& Astronomy, University College London, Gower Street, London, WC1E 6BT, UK}

\author[0000-0002-2890-3725]{Jamie E.\ Forero-Romero}
\affiliation{Departamento de F\'isica, Universidad de los Andes, Cra. 1 No. 18A-10, Edificio Ip, CP 111711, Bogot\'a, Colombia}
\affiliation{Observatorio Astron\'omico, Universidad de los Andes, Cra. 1 No. 18A-10, Edificio H, CP 111711 Bogot\'a, Colombia}

\author[0000-0003-3142-233X]{Satya Gontcho A Gontcho}
\affiliation{Lawrence Berkeley National Laboratory, 1 Cyclotron Road, Berkeley, CA 94720, USA}

\author{Klaus Honscheid}
\affiliation{Center for Cosmology and Astroparticle Physics, The Ohio State University, 191 West Woodruff Avenue, Columbus, OH 43210, USA}
\affiliation{Department of Physics, The Ohio State University, 191 West Woodruff Avenue, Columbus, OH 43210, USA}
\affiliation{The Ohio State University, Columbus, 43210 OH, USA}

\author[0000-0003-1838-8528]{Martin Landriau}
\affiliation{Lawrence Berkeley National Laboratory, 1 Cyclotron Road, Berkeley, CA 94720, USA}

\author[0000-0003-4962-8934]{Marc Manera}
\affiliation{Departament de F\'{i}sica, Serra H\'{u}nter, Universitat Aut\`{o}noma de Barcelona, 08193 Bellaterra (Barcelona), Spain}
\affiliation{Institut de F\'{i}sica d'Altes Energies (IFAE), The Barcelona Institute of Science and Technology, Campus UAB, 08193 Bellaterra Barcelona, Spain}

\author[0000-0002-4279-4182]{Paul Martini}
\affiliation{Department of Astronomy, The Ohio State University, 4055 McPherson Laboratory, 140 W 18th Avenue, Columbus, OH 43210, USA}
\affiliation{Center for Cosmology and AstroParticle Physics, The Ohio State University, 191 West Woodruff Avenue, Columbus, OH 43210, USA}
\affiliation{The Ohio State University, Columbus, 43210 OH, USA}

\author[0000-0002-1125-7384]{Aaron Meisner}
\affiliation{NSF's NOIRLab, 950 N. Cherry Ave., Tucson, AZ 85719, USA}
\author{Ramon Miquel}
\affiliation{Instituci\'{o} Catalana de Recerca i Estudis Avan\c{c}ats, Passeig de Llu\'{\i}s Companys, 23, 08010 Barcelona, Spain}
\affiliation{Institut de F\'{i}sica d'Altes Energies (IFAE), The Barcelona Institute of Science and Technology, Campus UAB, 08193 Bellaterra Barcelona, Spain}

\author[0000-0002-0644-5727]{Jundan Nie}
\affiliation{National Astronomical Observatories, Chinese Academy of Sciences, A20 Datun Rd., Chaoyang District, Beijing, 100012, P.R. China}

\author[0000-0002-0644-5727]{Will Percival}
\affiliation{Department of Physics and Astronomy, University of Waterloo, 200 University Ave W, Waterloo, ON N2L 3G1, Canada}
\affiliation{Perimeter Institute for Theoretical Physics, 31 Caroline St. North, Waterloo, ON N2L 2Y5, Canada}
\affiliation{Waterloo Centre for Astrophysics, University of Waterloo, 200 University Ave W, Waterloo, ON N2L 3G1, Canada}

\author[0000-0001-7145-8674]{Francisco Prada}
\affiliation{Instituto de Astrof\'{i}sica de Andaluc\'{i}a (CSIC), Glorieta de la Astronom\'{i}a, s/n, E-18008 Granada, Spain}

\author[0000-0001-5589-7116]{Mehdi Rezaie}
\affiliation{Department of Physics, Kansas State University, 116 Cardwell Hall, Manhattan, KS 66506, USA}

\author{Graziano Rossi}
\affiliation{Department of Physics and Astronomy, Sejong University, Seoul, 143-747, Korea}

\author[0000-0002-9646-8198]{Eusebio Sanchez}
\affiliation{CIEMAT, Avenida Complutense 40, E-28040 Madrid, Spain}

\author{Michael Schubnell}
\affiliation{Department of Physics, University of Michigan, Ann Arbor, MI 48109, USA}
\affiliation{University of Michigan, Ann Arbor, MI 48109, USA}

\author[0000-0003-1704-0781]{Gregory Tarl\'{e}}
\affiliation{University of Michigan, Ann Arbor, MI 48109, USA}

\author{Benjamin A.\ Weaver}
\affiliation{NSF's NOIRLab, 950 N. Cherry Ave., Tucson, AZ 85719, USA}

\author[0000-0002-4135-0977]{Zhimin Zhou}
\affiliation{National Astronomical Observatories, Chinese Academy of Sciences, A20 Datun Rd., Chaoyang District, Beijing, 100012, P.R. China}

\begin{abstract}
    We explore the galaxy-halo connection information that is available in low-redshift samples from the early data release of the Dark Energy Spectroscopic Instrument (DESI). We model the halo occupation distribution (HOD) from $z =$ 0.1-0.3 using Survey Validation 3 (SV3; a.k.a., the One-Percent Survey) data of the DESI Bright Galaxy Survey (BGS). In addition to more commonly used metrics, we incorporate counts-in-cylinders (CiC) measurements, which drastically tighten HOD constraints. Our analysis is aided by the Python package, \texttt{galtab}, which enables the rapid, precise prediction of CiC for any HOD model available in \texttt{halotools}. This methodology allows our Markov chains to converge with much fewer trial points, and enables even more drastic speedups due to its GPU portability. Our HOD fits constrain characteristic halo masses tightly and provide statistical evidence for assembly bias, especially at lower luminosity thresholds: the HOD of central galaxies in $z \sim 0.15$ samples with limiting absolute magnitude $M_r < -20.0$ and $M_r < -20.5$ samples is positively correlated with halo concentration with a significance of 99.9\% and 99.5\%, respectively. Our models also favor positive central assembly bias for the brighter $M_r < -21.0$ sample at $z\sim0.25$ (94.8\% significance), but there is no significant evidence for assembly bias with the same luminosity threshold at $z\sim0.15$. We provide our constraints for each threshold sample's characteristic halo masses, assembly bias, and other HOD parameters. These constraints are expected to be significantly tightened with future DESI data, which will span an area 100 times larger than that of SV3.
\end{abstract}

\section{Introduction}

The large-scale distribution of galaxies in the universe is a powerful probe of cosmological models \citep[e.g.,][]{Beutler:2011, Anderson:2012, Abbott:2018}. This is because galaxies trace the dark matter distribution, whose distribution is set by cosmological parameters and is well-characterized by modern simulations \citep[e.g.,][]{Klypin:2016, Ishiyama:2021}. However, for accurate cosmological inference, it is necessary to marginalize over the possible relationships between observational probes and the theoretical matter distribution. Therefore, leveraging large-scale structure to constrain cosmology requires flexible models of the galaxy-halo connection, and necessitates incorporating as much empirical information as possible to tightly constrain such flexible models.

Halos are thought to form central galaxies in their dense centers and accrete subhalos, which bring along their own central galaxies, becoming satellite galaxies of the primary halo. Therefore, the spatial clustering of most galaxy samples can be described well by a halo occupation distribution \citep[HOD; e.g.,][]{Berlind:Weinberg:2002, Zheng:2007}, which probabilistically connects the average number of central and satellite galaxies that a dark matter halo hosts to its mass. This formalism can be extended through additional parameters that lead to correlations between galaxy abundance and secondary halo properties \citep[i.e., galaxy assembly bias, e.g.,][]{Hearin:2016}, which can improve fit quality. As the data continues to improve, further variations to HOD models should be explored, e.g., by relaxing the assumption of a log-normal stellar-to-halo-mass relation or of a spatially isotropic Navarro-Frenk-White (NFW) distribution of satellite galaxies.

The most common observables used to constrain the galaxy-halo connection via spectroscopic galaxy samples are the number density and the projected two-point correlation function \wprp\ \citep[e.g.,][]{Zehavi:2005, Reddick:2013}. However, \citet{Wang:2019} has shown that the counts-in-cylinders (CiC) distribution $P(N_{\rm CiC})$ offers significant complementary information on the parameters of interest --- particularly those that control satellite occupation and assembly bias. As demonstrated by \citet{Storey-Fisher:2022}, it is also possible to quantify clustering information beyond the two-point function using the underdensity probability function and the density-marked correlation function. These studies highlight that even with existing datasets, incorporating different measurements of the large-scale structure can help optimize model fitting.

In this paper, we extend previous analyses by incorporating a novel spectroscopic dataset; implementing a new, more efficient CiC prediction framework; and demonstrating the gain these provide. We leverage data from the Dark Energy Spectroscopic Instrument \citep[DESI;][]{DESI:2022}, which will ultimately obtain spectroscopic redshifts of 40~million galaxies in an effort to precisely map the large-scale structure of a large volume of the observable universe. While the full dataset is still being collected, this work utilizes redshift measurements for more than 40,000 galaxies obtained by the Survey Validation 3 (SV3) component of the DESI early data release \citep{DESI:2023}.

We approximately adopt the best-fit flat-universe cosmology from \citet{Planck:2018}. The relevant cosmological parameters that we use are as follows: $h=0.6777$, $\Omega_{m,0}=0.30712$, $\Omega_{b,0}=0.048252$, and $T_{\rm CMB}=2.7255$~K. However, we scale all distance and distance-dependent values to units equivalent to setting the Hubble parameter to $h=1$ (e.g., $h^{-1}$Mpc).

This paper is organized as follows. We describe our data, model, and summary statistics in Section~\ref{sec:data}. We detail our novel methodologies for measuring and predicting CiC, through the \texttt{galtab} package, in Section~\ref{sec:cic}. We explain our parameter inference technique in Section~\ref{sec:constraints}, and discuss our conclusions in Section~\ref{sec:results}.

\section{Data}
\label{sec:data}

\subsection{DESI BGS}

\begin{figure*}[ht!]
    \centering
    \includegraphics[width=\textwidth]{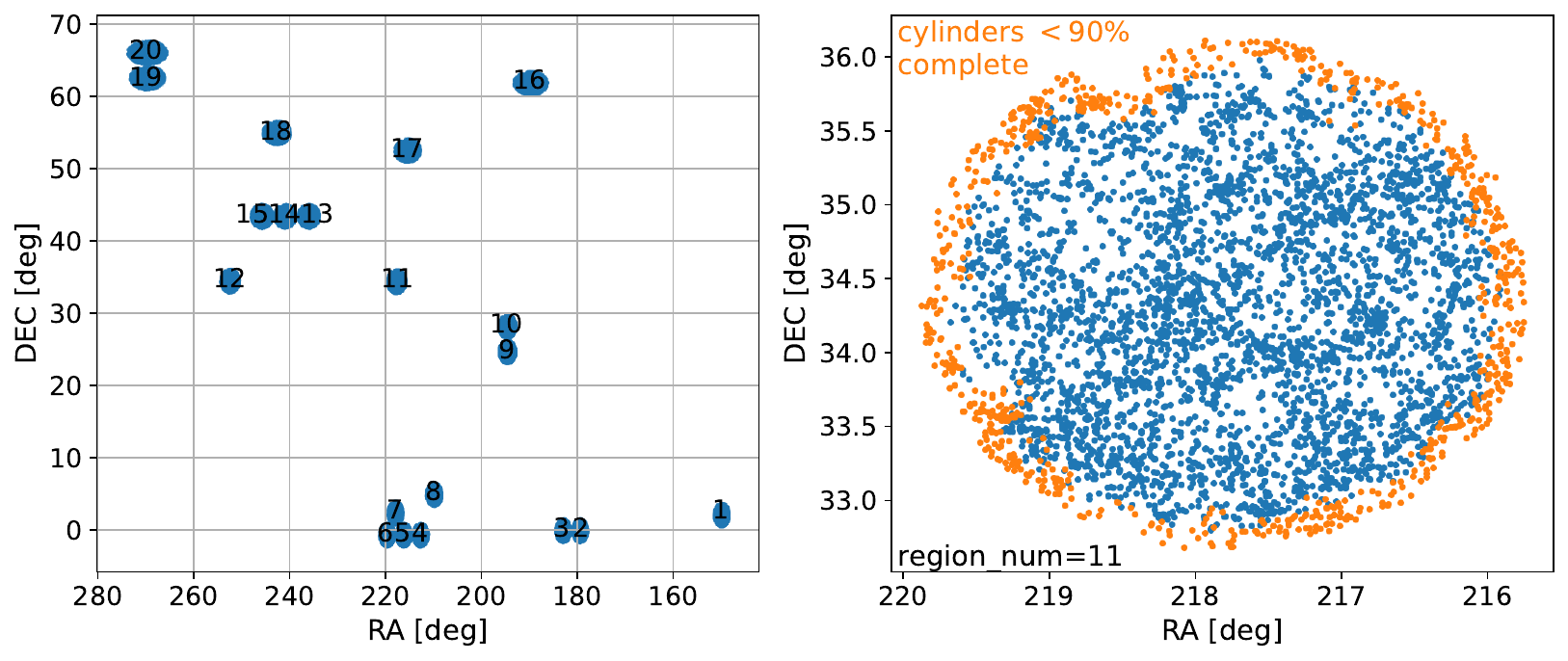}
    \caption{Footprint of the DESI Survey Validation 3 (SV3). The left panel displays the entire survey, broken up into twenty regions that are for the most part spatially isolated from each other. The right panel presents a close-up of the region labeled by the number 11 in the left panel. The points shown in orange, which are located primarily near the edge of the region, indicate objects excluded as cylinder centers in our CiC measurement, as described in Section~\ref{sec:iip-icp-weights}.}
    \label{fig:desi-footprint}
\end{figure*}

The DESI Bright Galaxy Survey (BGS) is a highly complete magnitude-limited spectroscopic survey of $z < 0.5$ galaxies, which aims to target galaxies over at least 14,000 square degrees down to a limit roughly two magnitudes fainter than the Sloan Digital Sky Survey \citep[SDSS;][]{SDSS:2009}. Our analyses only use the BGS Bright sample, which is complete down to an apparent r-band magnitude of $m_r < 19.5$. Because the DESI survey is still in progress at the time of this writing, we analyze only data from the Survey Validation 3 (SV3; \citealt{DESI:2023}) dataset (also known as the One-Percent Survey as it contains approximately 1\% of the anticipated volume of DESI). These data were obtained in over twenty sky regions totaling an area of 173.3 sq deg, as shown in Figure~\ref{fig:desi-footprint}. A significantly higher fraction of potential targets was observed in the SV3 fields than will be the case for typical DESI survey data due to the use of a denser tiling strategy, simplifying the corrections needed for our analysis.

We specifically use the SV3 Large Scale Structure (LSS) catalogs, which only include sources with secure spectroscopic redshift measurements, as described in \citet{DESI:2023}. These catalogs are well suited for clustering measurements since they are paired with 18 random realization files, each containing 2500 objects per deg$^2$ of sky coverage, and weights from 128 fiber assignment realizations. We also utilize $r$-band absolute magnitude measurements from \texttt{fastspecfit} (Moustakas et al.\ in prep.\footnote{\url{https://fastspecfit.readthedocs.io/}}), which are computed for an SDSS $r$-band response curve $K$-corrected to the $z=0.1$ reference frame using photometry from the Dark Energy Camera Legacy Survey \citep[DECaLS;][]{Dey:2019} and spectroscopic redshifts from DESI. Note that all references to absolute magnitudes in this paper, $M_r$, are scaled to $h=1$ units; therefore, they are equivalent to $M_r - 5\log h$ for all other values of the Hubble parameter.

\begin{figure}[ht!]
    \centering
    \includegraphics[width=0.47\textwidth]{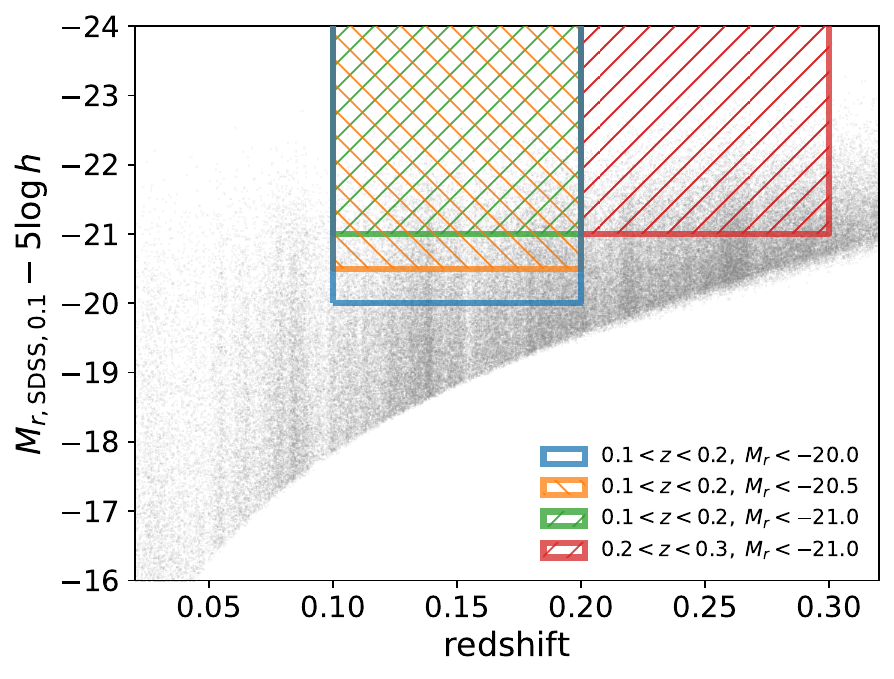}
    \caption{Distribution of $r$-band absolute magnitude $M_r$ vs.\ redshift. The full DESI BGS SV3 sample is shown by the grey points. Our four volume-limited, absolute magnitude-thresholded samples are constructed through the cuts represented by the corresponding colored boundaries.}
    \label{fig:rmag-vs-z-sample-cuts}
\end{figure}

We break this data into three volume-limited samples which each cover the redshift range $0.1 < z < 0.2$, constructed with absolute $r$-band absolute magnitude limits of $M_r < -20.0$, $-20.5$, and $-21.0$. We also define a fourth sample covering a slightly higher redshift range of $0.2 < z < 0.3$ with limit $M_r < -21.0$. We plot each sample cut in Figure~\ref{fig:rmag-vs-z-sample-cuts} and summarize these samples in Table~\ref{tab:sample-cuts}. Unless otherwise specified, all observational measurements in this paper are measured from one of these samples.

\startlongtable
\begin{deluxetable}{c|c|c|c}
\tablecaption{DESI subsamples used for our analyses. The full sample size is given by $N_{\rm tot}$, while $N_{\rm cyl}$ is the number of centers of the cylinders that meet our spatial completeness criteria. \label{tab:sample-cuts}}
\tablehead{
\colhead{$M_r$ threshold} & \colhead{Redshift range} & \colhead{$N_{\rm tot}$} & \colhead{$N_{\rm cyl}$}
}
\startdata
-20.0 & $0.1 < z < 0.2$ & 20,{}241 & 15,{}936\\
-20.5 & $0.1 < z < 0.2$ & 11,{}036 & 8,{}686\\
-21.0 & $0.1 < z < 0.2$ & 5,{}096 & 4,{}031\\
-21.0 & $0.2 < z < 0.3$ & 14,{}874 & 12,{}543\\
\enddata
\end{deluxetable}

\subsection{Model Galaxies}
\subsubsection{Small MultiDark Planck}

To study the galaxy-halo connection, we must compare DESI galaxy clustering data to an assumed distribution of underlying dark matter halos. For this halo distribution prior, we adopt the Small MultiDark Planck simulation \citep[SMDPL;][]{Klypin:2016}, which uses the same Planck cosmology that we assume in this work. This simulation was performed with $3840^3$ particles, but our analysis is based only upon the halo catalogs produced by applying the Rockstar halo finder \citep{Behroozi:2013}. We adopt the virial mass from Rockstar as our halo mass, $M_h$.

The particle mass of this simulation is roughly $10^8~M_\odot$, so all halos contributing to our analysis contain over $10^3$ particles. SMDPL covers a $400h^{-1}$~Mpc periodic cube, which is over ten times the volume of our SV3 samples. This is sufficiently large so that cosmic-variance-like uncertainties from the data dominate over the sample variance of this simulation volume.
However, future studies will need to use larger volume simulations to compensate for DESI's volume, which will be 100 times that of SV3.

\subsubsection{HOD Model}
\label{sec:hod}

We place model galaxies into the simulation cube by assuming that each halo hosts some number of central and satellite galaxies. To do this, we employ a decorated HOD model, where each central galaxy is placed at the center of its host halo, while the positions of satellite galaxies are drawn from their host halo's NFW profile. For a given halo, we assume that the number of central galaxies is drawn from a Bernoulli distribution, while the number of satellite galaxies is drawn from a Poisson distribution. In the standard \citet{Zheng:2007} HOD formalism, their means are functions of halo mass $M_h$ alone, described by

\begin{equation}
    \label{eq:hod-ncen}
    \langle N_{\rm cen} \rangle_{\rm std}(M_h) =  \frac{1}{2} \left(1 + {\rm erf}\left(\frac{\log(M_h / M_{\rm min})}{\sigma_{\log M}} \right) \right)
\end{equation}
and

\begin{equation}
    \label{eq:hod-nsat}
    \langle N_{\rm sat} \rangle_{\rm std}(M_h) = \left(\frac{M_h - M_0}{M_1}\right)^\alpha
\end{equation}
where $\log M_{\rm min}$, $\sigma_{\log M}$, $\alpha$, $\log M_1$, and $\log M_0$ are free parameters controlling the shape of the mean occupation functions. These parameters must be tuned separately for each magnitude threshold and redshift sample. We further parameterize $\log M_0$ into $Q_0$ using

\begin{equation}
    \log M_0 = \log M_{\rm min} + Q_0(\log M_1 - \log M_{\rm min})
\end{equation}
which helps us ensure that $\log M_0$ always stays between $\log M_{\rm min}$ and $\log M_1$ to preserve its sensitivity to, and the stability of, our summary statistics.

Adding further flexibility into our model, we include assembly bias parameters $A_{\rm cen}$ and $A_{\rm sat}$ to introduce a halo occupation dependence on the NFW concentration. By definition, each of these parameters can only range from [-1, 1], and allow for the redistribution of central and satellite occupation, respectively, from low to high concentration halos for positive $A$, or vice versa. Following \cite{Zentner:2019}, we modify the halo occupations according to a perturbation $\delta N_{\rm gal}$, with a sign dependent on if a halo is in the upper or lower 50-percentile of concentration ($c_{\rm high}$ or $c_{\rm low}$) in a narrow mass bin, such that

\begin{equation}
    \langle N_{\rm gal} \rangle(M_h, c_{\rm high}) = \langle N_{\rm gal} \rangle_{\rm std} + \delta N_{\rm gal}
\end{equation}
\begin{equation}
    \langle N_{\rm gal} \rangle(M_h, c_{\rm low}) = \langle N_{\rm gal} \rangle_{\rm std} - \delta N_{\rm gal}
\end{equation}
where
\begin{equation}
    \delta N_{\rm sat} = A_{\rm sat}\langle N_{\rm sat} \rangle_{\rm std}
\end{equation}
\begin{equation}
    \delta N_{\rm cen} = A_{\rm cen} \left( 0.5 - \left\lvert 0.5 - \langle N_{\rm cen} \rangle_{\rm std} \right\rvert \right)
\end{equation}

For all samples, we adopt uniform priors on our model parameters with the following bounds: $\log M_{\rm min} \in [9, 16]$; $\sigma_{\log M} \in [10^{-5}, 5]$; $\log M_1 \in [10, 16]$; $Q_0 \in [0, 1]$; $\alpha \in [10^{-5}, 5]$; $A_{\rm cen} \in [-1, 1]$; and $A_{\rm sat} \in [-1, 1]$. These bounds are very wide compared to our data constraints, so they do not strongly influence our fits, except for the upper limit on $A_{\rm cen}$.

\subsection{Summary Statistics}
\label{sec:summary-stats}

To extract clustering information from each galaxy sample, we use three summary statistics: number density $n_{\rm gal}$, the projected two-point correlation function \wprp, and the CiC distribution $P(N_{\rm CiC})$. We seek a good agreement of these summary statistics, as measured in the model galaxies vs.\ the DESI data, to validate our model. We display our best-fit models against the corresponding observations of these three summary statistics for each sample in Figure~\ref{fig:summary-stats}.

The number density is calculated via the sum of the inverse individual probability (IIP; see Section~\ref{sec:iip-icp-weights}) weights of the galaxies in the sample divided by the comoving volume they were sampled from.
For the HOD number density predictions, the comoving volume of SMDPL is $400^3h^{-3}{\rm Mpc}^3$, while the volumes of the DESI samples depend on the redshift cuts and the survey area. The DESI SV3 BGS survey area is 173.3 sq~deg, which corresponds to comoving volumes of $2.83 \times 10^6h^{-3}{\rm Mpc}^3$ and $6.95 \times 10^6h^{-3}{\rm Mpc}^3$ for samples with redshift ranges of $0.1 < z < 0.2$ and $0.2 < z < 0.3$, respectively.

The projected two-point correlation function is a common way to quantify spatial clustering in redshift space at various physical scales. By integrating over the line-of-sight dimension, this statistic decreases the dependence of the inferred clustering on redshift-space distortions. It is defined by

\begin{equation}
    w_{\rm p}(r_{\rm p}) = 2 \int_0^{\pi_{\rm max}} \xi(r_{\rm p}, \pi) d\pi
\end{equation}
where $\xi$ is the two-point correlation function, $\pi$ is line-of-sight separation distance, and $r_{\rm p}$ is perpendicular separation distance. For consistency with \cite{Wang:2022}, we choose $\pi_{\rm max} = 40h^{-1}{\rm Mpc}$ and use twelve logarithmically spaced bins between $r_{\rm p}$ of $0.158h^{-1}{\rm Mpc}$ and $39.81h^{-1}{\rm Mpc}$. We concatenate all 18 random files from the SV3 LSS catalogs but draw a random 20\% subsample which is sufficient so that the randoms contribute negligibly to our uncertainties. We utilize the \texttt{pycorr}\footnote{\url{https://github.com/cosmodesi/pycorr}} package to apply the \citet{Landy:Szalay:1993} estimator, line-of-sight integration, and fiber assignment weights. The performance-critical pair searching is powered by \texttt{Corrfunc} \citep{Sinha:2020}.

We choose to use projected statistics, which integrate over line-of-sight separations up to $40h^{-1}{\rm Mpc}$ for \wprp\ and $10h^{-1}{\rm Mpc}$ for CiC. Since the line-of-sight position is significantly more uncertain (due to peculiar velocities) than the other two coordinates, there is more clustering information per degree of freedom when the line-of-sight direction is discretized into broader bins. However, it is possible to summarize the clustering information even more thoroughly by replacing \wprp\ with the monopole and quadrupole of the redshift-space two-point correlation function, at the expense of additional degrees of freedom.

\begin{figure*}[ht!]
    \centering
    \includegraphics[width=\textwidth]{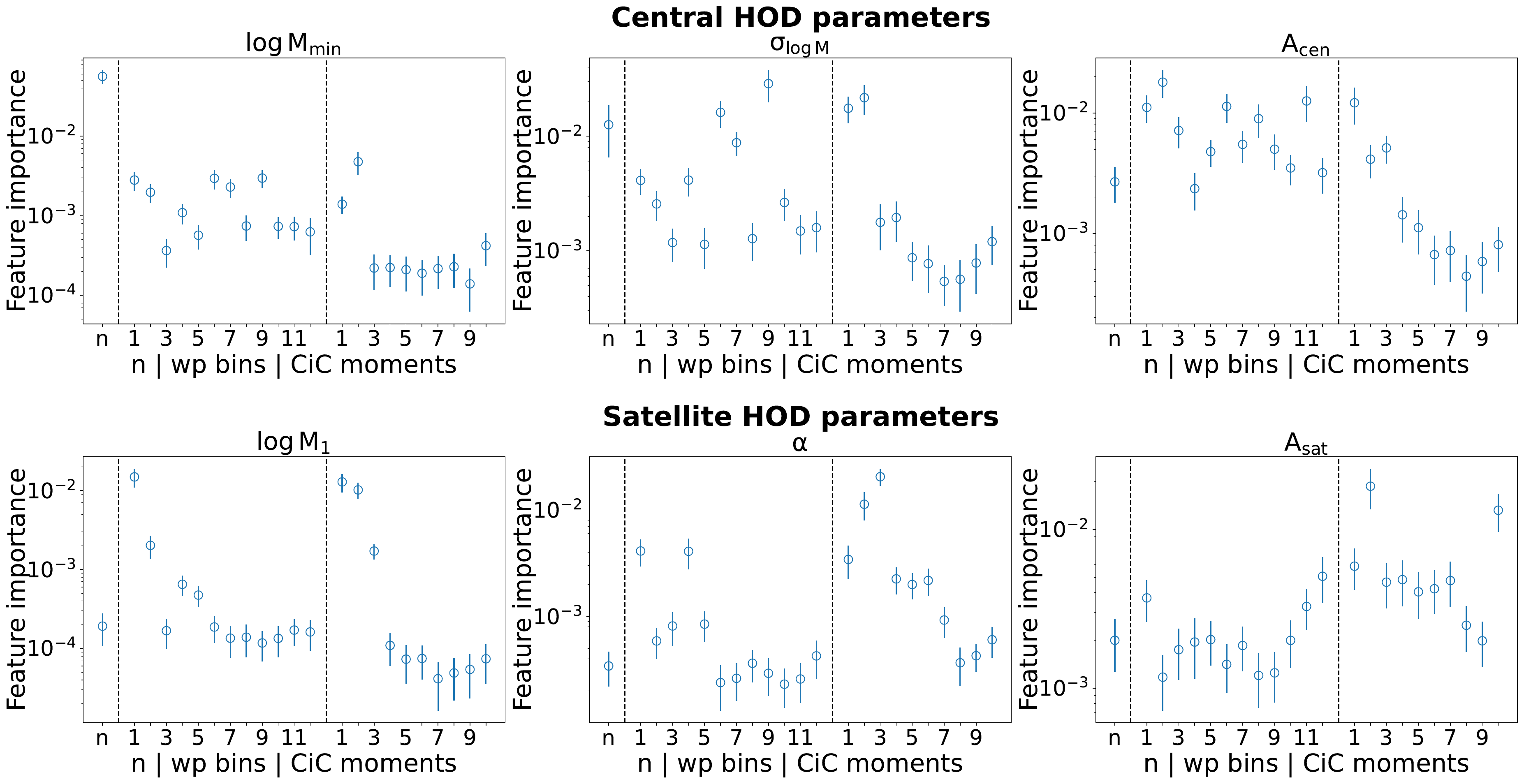}
    \caption{
    SHAP feature importances for each of our summary statistics for inferring HOD parameters. Each panel plots the importance of each feature (i.e., each quantity that is used to predict the HOD parameters via a machine learning model), calculated by the mean absolute SHAP value for the given HOD parameter.  Summary statistics with high feature importance are more useful for predicting the parameter. For the satellite HOD parameters (bottom row), the first few CiC moments provide the majority of the constraining information. See Figure~\ref{fig:shap-beeswarm} for beeswarm plots of the full distribution of SHAP values of the six most important features for each parameter.}
    \label{fig:feature-importance}
\end{figure*}

Counts-in-cylinders (CiC) is a type of counts-in-cells statistic \citep[i.e., it quantifies the local density of points in a cell of a given geometry; the development of such metrics has a long history; e.g.,][]{Hubble:1936, Zwicky:1957, White:1979, Adelberger:1998} that defines neighbors using a cylindrical cell along the line-of-sight direction. As done by \cite{Wang:2022}, we center a cylinder of radius $R_{\rm CiC} = 2h^{-1}{\rm Mpc}$ and half-length $L_{\rm CiC} = 10h^{-1}{\rm Mpc}$ around each galaxy in the sample. We count the number of near neighbors there are around each galaxy, $N_{\rm CiC}$, enclosed by this cylindrical cell, excluding self-counting so that $N_{\rm CiC}=0$ is possible. Cylinders of this scale primarily probe the number of intra-halo galaxies and are therefore sensitive to satellite occupation as well as assembly bias \citep[see][and references within]{Wang:2019}. Conveniently, using a small cylindrical volume is also a computationally favorable choice. In principle, further information could be obtained by simultaneously fitting several different cylinder sizes (i.e., varying $L_{\rm CiC}$ or $R_{\rm CiC}$), although this is unlikely to be worth the extra computational expense. We evaluate the CiC distribution $P(N_{\rm CiC})$ in bins of $N_{\rm CiC}$, for which we use ten linearly spaced bins bounded by $-0.5$ and 9.5 plus twenty logarithmically spaced bins between 9.5 and 149.5. Alternatively, if the runtime or covariance matrix dimensionality is a major concern, the majority of available information can be captured by computing the first three to five moments of the $N_{\rm CiC}$ distribution (importance shown in Figure~\ref{fig:feature-importance}). We describe our methods used to compute counts-in-cylinders in detail in Section~\ref{sec:cic}.

We test the ability of each summary statistic to inform the HOD by sampling uniformly from HOD parameters around their 1$\sigma$ confidence interval from the \citet{Wang:2022} $M_r < -20.5$ sample. See Appendix~\ref{sec:appendix-shap} for a detailed discussion of this procedure. In brief, we predict each of our summary statistics including the first ten CiC moments. We then train a random forest \citep{Breiman:2001} to predict the HOD parameters from these summary statistics and provide a visualization of the resulting SHapley Additive exPlanations \citep[SHAP;][]{Lundberg:2017} feature importance in Figure~\ref{fig:feature-importance}. We can conclude that number density is highly important for predicting $\log M_{\rm min}$, the two-point correlation function is broadly informative across all parameters, and the first few CiC moments are particularly important for constraining satellite HOD parameters.

\subsubsection{Covariance of Summary Statistics}
\label{sec:cov}

To constrain our HOD model, we compare the following summary statistics as measured in our data to model predictions: number density; the two-point correlation function (computed in 12 bins in $r_{\rm p}$); and CiC (for 28 bins in $N_{\rm CiC}$). We calculate the covariance matrix of these summary statistics by jackknife resampling using the 20 regions displayed in Figure~\ref{fig:desi-footprint}.

To do this, we perform a measurement of every summary statistic simultaneously on the subset of data that includes all but one jackknife region. We repeat this process for each combination of 19 jackknife regions to obtain $N_{\rm J}=20$ jackknife realizations. The covariance matrix of our summary statistics can then be estimated by

\begin{equation}
    \label{eq:cov-jackknife}
    \Sigma_{ij} = \frac{N_{\rm J} - 1}{N_{\rm J}} \sum_{k=0}^{N_{\rm J}}(x_{ik} - \bar{x}_i)(x_{jk} - \bar{x}_j)
\end{equation}
where $\bar{x}_i$ is the $i$th summary statistic measured in the entire dataset, and $x_{ik}$ is the $i$th summary statistic measured in the $k$th jackknife realization.

Using only 20 jackknife regions for this purpose is a somewhat noisy estimator of the covariance matrix. However, breaking them into even smaller regions would severely violate the assumption that the jackknife regions are independent of each other. Note that calculating CiC in small regions is particularly problematic because data near the edges must be removed.

\section{Counts-in-Cylinders}
\label{sec:cic}

Counts-in-cylinders (CiC; derived in \citealt{Peebles:1980} and previously used by \citealt{Reid:Spergel:2009, Wang:2022}) is sensitive to higher-order n-point functions, which makes it complementary to two-point statistics commonly used in the literature. Despite its utility, CiC is not widely adopted in galaxy-halo connection studies, due to difficulties in correcting for systematics, excessive computational time, and significantly increased dimensionality of the full covariance matrix. In this section, we present our methodology to mitigate all of these problems and implement each of these methods in an open-source Python package \texttt{galtab}\footnote{\href{https://github.com/AlanPearl/galtab}{https://github.com/AlanPearl/galtab}}.

After a brief explanation of our observational cylinder geometry in Section~\ref{sec:obs-cic-geometry}, we present our weighting method in Section~\ref{sec:iip-icp-weights} based on individual inverse probabilities and inverse conditional probabilities (IIP\texttimes ICP), which corrects CiC calculations to account for clustering bias in surveys with fiber collisions. This approach is analogous to and inspired by pair inverse probabilities (PIP) weighting \citep{Bianchi:Percival:2017}, which we used to correct our \wprp\ measurement. To minimize the dimensionality of the covariance matrix, we suggest using only the first three to five moments of the CiC distribution, defined in Section~\ref{sec:moments}, which retain most of the constraining information. Our analysis uses information from the entire CiC distribution, but the constraining power should not be significantly diminished by using only the first five CiC moments instead.

Additionally, we present a galaxy placeholder pretabulation method in Section~\ref{sec:pretab} to speed up our Markov-chain Monte Carlo (MCMC) procedure.
This makes our CiC prediction runtime comparable to traditional Monte Carlo \wprp\ prediction methods but with the significant advantage of producing precise, deterministic values, which yield much higher MCMC sampling efficiency than stochastic Monte Carlo predictions. In Sections~\ref{sec:monte-carlo-prediction} and~\ref{sec:analytic-prediction}, we present two different CiC prediction frameworks and discuss their respective use cases.

\subsection{Observational Cylinder Geometry}
\label{sec:obs-cic-geometry}

While a cylinder perfectly aligns with the velocity distortion in an idealized simulation, for observations, we must slightly distort its curved rectangular face into a truncated cone so that it is parallel to the line-of-sight direction (like a light cone). We also allow a slight curve to this truncated cone's top and bottom circular faces, so they are normal to the line of sight. Therefore, we only have two search criteria in our CiC search: maximum angular and line-of-sight separations. The line-of-sight separation cut is $L_{\rm CiC}$ and we define the angular separation cut to be

\begin{equation}
    \theta_{\rm CiC} = \arccos{\left(1 - \frac{3 R_{\rm CiC}^2 L_{\rm CiC}}{(d + L_{\rm CiC})^3 - (d - L_{\rm CiC})^3}\right)}
\end{equation}
where $d$ is the comoving distance to the galaxy centered by the ``cylinder''. This ensures that its volume is still precisely $2 \pi R_{\rm CiC}^2 L_{\rm CiC}$, and $\theta_{\rm CiC} \approx R_{\rm CiC}/d$ as $d \to \infty$.

\subsection{IIP\texttimes ICP Weighting}
\label{sec:iip-icp-weights}

In order to account for fiber collisions, the DESI Large-Scale Structure catalogs come with ``bitweights'' columns. These columns represent 128 true (1) or false (0) values for each object corresponding to 128 fiber assignment realizations stored as a bitmask. To make the realizations independent of one another, the targets are randomly assigned sub-priority values, and the survey footprint is slightly dithered, following the methods outlined in \citet{Smith:2019}. The true fiber assignments for the One-Percent Survey are effectively a 129th realization in which all data in our sample have an understood simultaneous true value. Therefore, the probability of assigning a fiber to the $i$th galaxy is

\begin{equation}
    \label{eq:prob-i}
    P(i) = \frac{\texttt{sum(bitweights[i])} + 1}{129},
\end{equation}
while the probability of simultaneously assigning fibers to both the $i$th and $j$th galaxies is

\begin{equation}
    \label{eq:prob-i-and-j}
    P(i\ \&\ j) = \frac{\texttt{sum(bitweights[i] \& bitweights[j])} + 1}{129},
\end{equation}
where \texttt{sum} and \texttt{\&} are bitwise operations. Thanks to the high fiber completeness of SV3, the average value of $P(i)$ is 0.984.

In order to measure the CiC distribution, we must calculate the expectation value of the number of galaxies we expect to find in the cylinder around every galaxy individually, $N_{{\rm CiC}, i}$. For this task, we sum the inverse conditional probabilities (ICPs) of each neighboring galaxy's fiber assignment (conditional on the fiber assignment of the cylinder's central galaxy). Using the definitions from Equations~\ref{eq:prob-i} and~\ref{eq:prob-i-and-j},

\begin{equation}
    {\rm ICP}_{j|i} = \frac{P(i)}{P(i\ \&\ j)}
\end{equation}
\begin{equation}
    N_{{\rm CiC}, i} = \frac{1}{f_{\rm rand}}\sum_{j \in C_i} {\rm ICP}_{j|i}
\end{equation}
where $C_i$ is the set of indices of galaxies contained by the cylinder surrounding the $i$th galaxy (excluding the $i$th galaxy itself), and $f_{\rm rand}$ is the number of randoms enclosed in its cylinder's angular selection divided by the expected number occupying a circle of angular radius $\theta_{\rm CiC}$, which accounts for incompleteness in footprint coverage. Note that we do not include cylinders with $f_{\rm rand} < 0.9$. This cut excludes approximately 21\% of the cylinders at $z\sim0.15$ and 16\% of the cylinders at $z\sim0.25$, as listed in Table~\ref{tab:sample-cuts}. The excluded galaxies are primarily located near the edge of the footprint, as seen in Figure~\ref{fig:desi-footprint}.

We measure $P(N_{\rm CiC})$ from the sample distribution of $N_{{\rm CiC}, i}$ values, but we need to overweight objects in dense regions of the sky that have been undersampled. Therefore, we weight objects by their inverse individual probability (IIP). The IIP of the $i$th galaxy is simply

\begin{equation}
    {\rm IIP}_i = \frac{1}{P(i)}.
\end{equation}

Finally, for our binned histogram measurements of $P(N_{\rm CiC})$, we split each ${\rm IIP}_i$ into two parts, ${\rm IIP}_{i+}$ and ${\rm IIP}_{i-}$. These weights are applied to the integers above and below $N_{{\rm CiC}, i}$, respectively, and are proportional to one minus that integer's distance from $N_{{\rm CiC}, i}$ so that

\begin{equation}
    \label{eq:iip-split}
    \frac{{\rm IIP}_{i+} \lceil N_{{\rm CiC}, i} \rceil + {\rm IIP}_{i-} \lfloor N_{{\rm CiC}, i} \rfloor}{{\rm IIP}_i} = N_{{\rm CiC}, i}.
\end{equation}
This allows us to only assign histogram values to integer cylinder counts ($\lceil N_{{\rm CiC}, i} \rceil$ and $\lfloor N_{{\rm CiC}, i} \rfloor$), even though $N_{{\rm CiC}, i}$ is not necessarily an integer. This step is necessary because $P(N_{\rm CiC})$ is formally a probability mass function, not a probability distribution.

\subsection{Calculating the CiC Moments}
\label{sec:moments}

In order to decrease the dimensionality of the covariance matrix, one may choose to condense the information contained in the CiC distribution into its first few moments, which we define as

\begin{equation}
    \label{eq:moment-1}
    \mu_1 = \sum_{i=1}^N w_i N_{{\rm CiC}, i}
\end{equation}
\begin{equation}
    \label{eq:moment-2}
    \mu_2 = \sqrt{\sum_{i=1}^N w_i (N_{{\rm CiC}, i} - \mu_1)^2}
\end{equation}
\begin{equation}
    \label{eq:moment-k}
    \mu_{k>2} = \frac{1}{\mu_2^k}\sum_{i=1}^N w_i (N_{{\rm CiC}, i} - \mu_1)^k.
\end{equation}
where $N_{{\rm CiC}, i}$ (in practice, this is split into $\lceil N_{{\rm CiC}, i} \rceil$ and $\lfloor N_{{\rm CiC}, i} \rfloor$; see Section~\ref{sec:iip-icp-weights}) is the number of neighbors inside the cylinder surrounding a galaxy in the sample and $w_i$ is the corresponding IIP weight, but normalized to $\sum w_i = 1$. Note that $\mu_1$ is the mean, $\mu_2$ is the standard deviation, and for $k > 2$, $\mu_k$ are standardized central moments (skewness, kurtosis, etc.), uncorrected for degree-of-freedom bias, which is a negligible source of systematics for large sample sizes compared to other uncertainties. In some figures, we refer to $\mu_k$ as CiC$_k$ to be explicit that they are moments of CiC.

\subsection{Pretabulation with Placeholder Galaxies}
\label{sec:pretab}

\begin{figure*}[ht!]
    \centering
    \includegraphics[width=\textwidth]{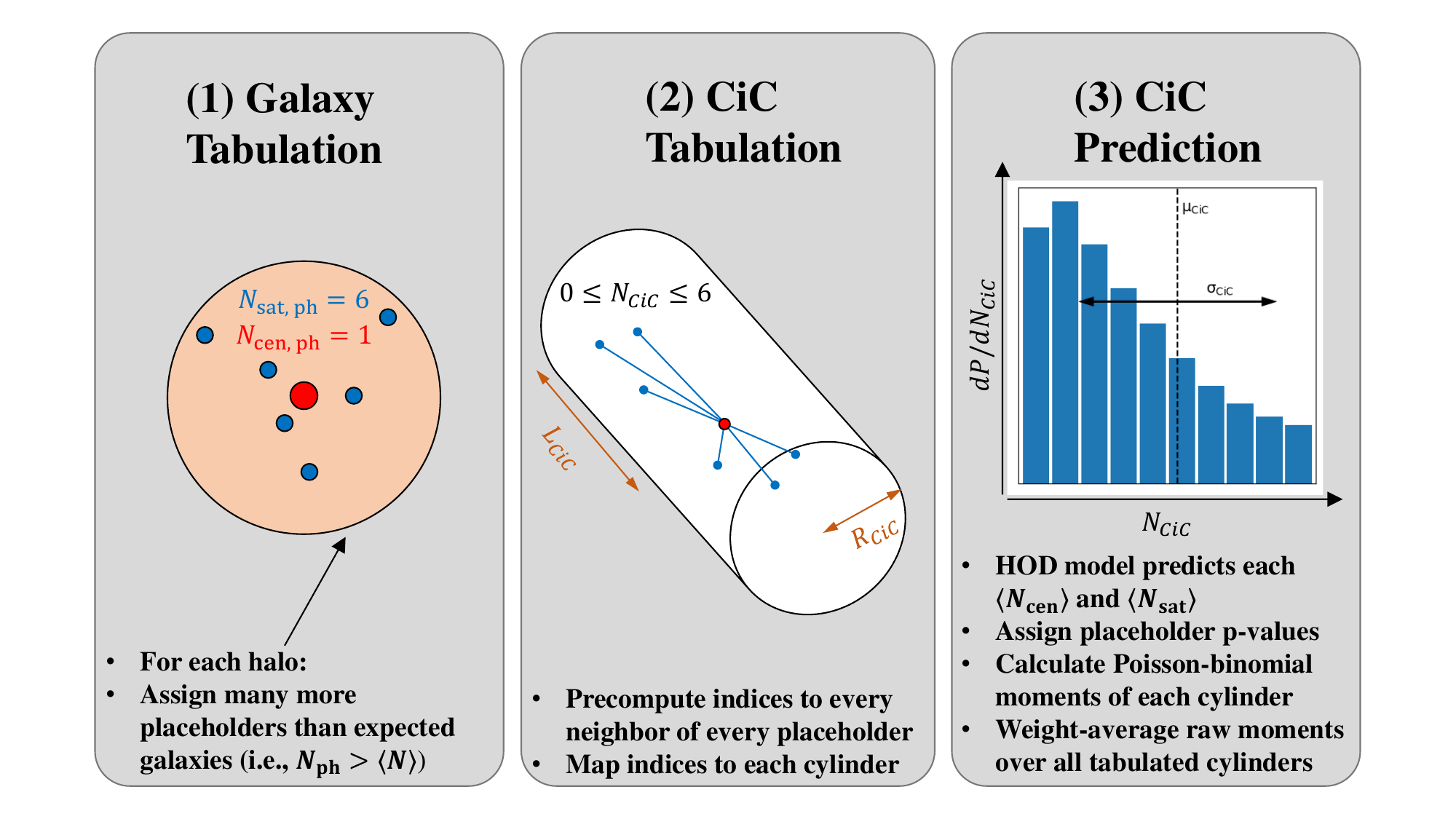}
    \caption{Demonstration of our placeholder algorithm used to pretabulate counts-in-cylinders pair indices. Given a fiducial model, we populate placeholder centrals for most halos with a non-zero probability of hosting a halo. We populate many more placeholder satellites than expected in the fiducial model so that the resulting binomial satellite occupation distribution sufficiently resembles the assumed Poisson distribution. We then tabulate the placeholder indices in each halo for rapid CiC prediction using one of the two modes described in Sections~\ref{sec:monte-carlo-prediction} and~\ref{sec:analytic-prediction}.}
    \label{fig:placeholder-cartoon}
\end{figure*}

Predictions of CiC from Monte Carlo HOD realizations are notoriously slow and noisy. This stochasticity reduces the sampling efficiency of Monte Carlo explorations of model parameter space by decreasing the acceptance rate which, in turn, increases the autocorrelation length of MCMC chains and necessitates longer chains and run times. To remedy this, we have developed a method to calculate precise, deterministic CiC predictions by pretabulating placeholder galaxies inside simulated halos.

Our procedure, illustrated in Figure~\ref{fig:placeholder-cartoon}, requires a fiducial HOD model to compute the expected occupation, $\langle N_{\rm cen} \rangle$ and $\langle N_{\rm sat} \rangle$, for each halo. For our fiducial model, we choose the best fit of \citet{Wang:2022} that corresponds to the magnitude threshold of each of our samples. We populate each halo with $N_{\rm cen, ph}$ central placeholders and $N_{\rm sat, ph}$ satellite placeholders. We determine the number of satellite placeholders for each halo with the hyperparameter $W_{\rm max}$ according to the equation

\begin{equation}
    N_{\rm sat, ph} = \left\lceil \frac{\langle N_{\rm sat} \rangle}{W_{\rm max}} \right\rceil
\end{equation}
which ensures that, for fiducial model predictions, there are enough satellite placeholders that their individual weights are less than or equal to $W_{\rm max}$.

For centrals, we define a hyperparameter $Q_{\rm min}$ that sets the minimum quantile of central galaxies for which to populate a central placeholder. In practice, we set $N_{\rm cen, ph}=1$ for all halos with $\langle N_{\rm cen} \rangle \geq \langle N_{\rm cen} \rangle_{\rm min}$, and $N_{\rm cen, ph}=0$ otherwise. To solve for $\langle N_{\rm cen} \rangle_{\rm min}$, we numerically integrate and invert

\begin{equation}
    Q_{\rm min} = \frac{\int_{\langle N_{\rm cen} \rangle_{\rm min}}^1 \Phi(\langle N_{\rm cen} \rangle)\langle N_{\rm cen} \rangle d\langle N_{\rm cen} \rangle}{\int_0^1 \Phi(\langle N_{\rm cen} \rangle)\langle N_{\rm cen} \rangle d\langle N_{\rm cen} \rangle}
\end{equation}
where $\Phi(\langle N_{\rm cen} \rangle) d\langle N_{\rm cen} \rangle$ is the number density of halos with expected central occupation between $\langle N_{\rm cen} \rangle$ and $\langle N_{\rm cen} \rangle + d\langle N_{\rm cen} \rangle$. This essentially places a minimum halo mass that varies for our HOD samples, ranging from $\sim$1-3$\times10^{11} M_\odot$.

To balance accuracy and runtime (see Figure~\ref{fig:accuracy-runtime}), we set $W_{\rm max}=0.05$ and $Q_{\rm min}=10^{-4}$. In \texttt{galtab}, these hyperparameters can be tuned via the \texttt{max\_weight} and \texttt{min\_quant} keyword arguments, respectively. After tabulation, one may choose any parameters for the HOD model and obtain a new prediction of $\langle N_{\rm X} \rangle$ for each halo and for each galaxy type denoted by X: central or satellite. Each placeholder galaxy is then assigned a probability value, $P_i=\langle N_{\rm X} \rangle / N_{\rm X, ph}$. 

As is usually done in Monte Carlo HOD realizations, these galaxy probability values are assumed to be independent. Therefore, the halo occupation of centrals follows a Bernoulli distribution, the same as typical Monte Carlo frameworks. However, the halo occupation of satellites follows a binomial distribution in our framework, which only converges to the desired Poisson distribution in the low $P_i \lesssim 0.05$ limit, hence our choice of $W_{\rm max}=0.05$.

Finally, a single counts-in-cylinder search is required (we use the \texttt{halotools} implementation for this) to obtain a list of the indices of possible neighbors for each placeholder. This allows us to rapidly calculate our CiC metric, as described in the following sections.

\subsection{Pretabulated CiC Prediction: Monte Carlo Mode}
\label{sec:monte-carlo-prediction}

In order to calculate the CiC distribution $P(N_{\rm CiC})$ from the probability values of our pretabulated galaxies, we must consider the probability of each possible value of $N_{{\rm CiC}, i}$ for each cylinder $i$. The full CiC distribution is simply the weighted superposition of that of each cylinder. We write this as

\begin{equation}
    \label{eq:cic-dist-superposition}
    P(N_{\rm CiC}) = \frac{\sum\limits_{i=1}^{N} P_i P(N_{{\rm CiC}, i})}{\sum\limits_{i=1}^N P_i}.
\end{equation}

In general, each $P(N_{{\rm CiC}, i})$ is a Poisson binomial distribution, whose exact calculation scales exponentially with the number of neighbors in the $i$th cylinder, which is infeasible. Therefore, we approximate this distribution for each cylinder using a Monte Carlo method. To do this, we pretabulate $n_{\rm MC}$ random seeds over $[0, 1)$ for each galaxy, which we use as Bernoulli quantiles after assigning the $P_i$ of each placeholder.
This effectively creates $n_{\rm MC}$ independent realizations that can produce quasi-deterministic and almost continuous (but non-differentiable) predictions of the boolean values that decide whether a given placeholder is populated. We construct each $P(N_{{\rm CiC}, i})$ as a histogram of the number of populated neighbors drawn by the $n_{\rm MC}$ realizations. We find that $n_{\rm MC}=10$ produces reasonably stable results without excessive runtime.

As an alternative to the Monte Carlo mode predictions described in this section, we have also implemented analytic mode predictions, which we will describe in Section~\ref{sec:analytic-prediction}. The analytic mode can predict CiC moments more efficiently, without invoking random seeds, allowing them to be perfectly continuous and differentiable. Therefore, when predicting CiC moments, it is recommended to use the analytic mode described in Section~\ref{sec:analytic-prediction} (and this is the default functionality for CiC moment prediction) instead of the Monte Carlo mode.

\subsection{Pretabulated CiC Prediction: Analytic Mode}
\label{sec:analytic-prediction}

Although the full $P(N_{\rm CiC})$ distribution cannot be calculated analytically from our galaxy placeholders, the moments of this distribution can. As a simple example, the mean of this distribution is simply the weighted average of the individual means

\begin{equation}
    \label{eq:mean-from-means}
    \langle N_{\rm CiC} \rangle = \frac{\sum\limits_{i=1}^{N} P_i \langle N_{{\rm CiC}, i} \rangle}{\sum\limits_{i=1}^N P_i}
\end{equation}
where
\begin{equation}
\label{eq:means-from-probs}
    \langle N_{{\rm CiC}, i} \rangle = \sum_{j \in C_i} P_j
\end{equation}
and $C_i$ is the set of indices of galaxies contained by the cylinder surrounding, but not including, the $i$th galaxy.

It is possible to calculate a similar relation for the standard deviation and the higher standardized moments we have defined in Equations~\ref{eq:moment-2} and~\ref{eq:moment-k}. However, these relations are much more complicated. Note that the mean is a special case because it is the first raw moment (which allows Equation~\ref{eq:mean-from-means}) as well as the first cumulant (which allows Equation~\ref{eq:means-from-probs}).

Cumulants are a type of moment that have a special property that they are additive for random variables which are the sum of other random variables. For example, the number of neighbors in the $i$th cylinder is a random variable, which is the sum of the occupation of each of its pretabulated placeholder companions, which themselves are random variables:

\begin{equation}
    \label{eq:ncic-sum-of-bernoulli-vars}
    N_{{\rm CiC}, i} = \sum_{j \in C_i} X_j
\end{equation}
where $X_j$ is the occupation of the $j$th placeholder, which follows a Bernoulli distribution (0 or 1) with mean $P_j$. Therefore, the first cumulant of this Bernoulli distribution is $\kappa_1(X_j) = P_j$, and the subsequent Bernoulli cumulants can be derived from the recursion relation

\begin{equation}
    \kappa_{k+1}(X_j) = P_j (1 - P_j) \frac{d\kappa_k(X_j)}{dP_j}.
\end{equation}

Given the first $k_{\rm max}$ Bernoulli cumulants of each placeholder in $C_i$, we can calculate the first $k_{\rm max}$ Poisson binomial cumulants of the $i$th cylinder. We simply take the $k$th cumulant of each random variable on both sides of Equation~\ref{eq:ncic-sum-of-bernoulli-vars}:

\begin{equation}
    \label{eq:cumulant-sum}
    \kappa_{k}(N_{{\rm CiC}, i}) = \sum_{j \in C_i} \kappa_k(X_j).
\end{equation}

From the moments of each $N_{{\rm CiC}, i}$, we would like the moments of the combined CiC distribution, which is a weighted superposition of each individual cylinder's distribution, as expressed in Equation~\ref{eq:cic-dist-superposition}. For this step, the most convenient set of moments to use are raw moments. The $k$th raw moment of $N_{{\rm CiC}, i}$ can be obtained from its first $k$ cumulants according to

\begin{equation}
    \langle N_{{\rm CiC}, i}^k \rangle = \kappa_k(N_{{\rm CiC}, i}) + \sum\limits_{j=1}^{k-1} \kappa_j(N_{{\rm CiC}, i}) \langle N_{{\rm CiC}, i}^{k-j} \rangle.
\end{equation}

From these individual $k$th raw moments, we can calculate the $k$th raw moment of their superposition using a simple weighted average:

\begin{equation}
    \langle N_{\rm CiC}^k \rangle = \frac{\sum\limits_{i=1}^{N} P_i \langle N_{{\rm CiC}, i}^k \rangle}{\sum\limits_{i=1}^{N} P_i}.
\end{equation}

The first raw moment is $\mu_1$, but the remaining $\mu_k$ for $2 \leq k \leq k_{\rm max}$ depend on central moments. Therefore, the final nontrivial step of our analytic prediction framework is to calculate the central moments using the following binomial expansion:

\begin{equation}
    \langle (N_{\rm CiC} - \langle N_{\rm CiC}\rangle)^k \rangle = \sum\limits_{j=0}^k {k\choose j} (-1)^{k-j} \langle N_{\rm CiC}^j \rangle \langle N_{\rm CiC} \rangle^{k-j}
\end{equation}
from which we can calculate the standard moments given in Equations~\ref{eq:moment-1} through~\ref{eq:moment-k} using

\begin{equation}
    \mu_1 = \langle N_{\rm CiC} \rangle,
\end{equation}

\begin{equation}
    \mu_2 = \sqrt{\langle (N_{\rm CiC} - \langle N_{\rm CiC} \rangle)^2 \rangle},
\end{equation}
and

\begin{equation}
    \mu_{k>2} = \frac{1}{\mu_2^k} \langle(N_{\rm CiC} - \langle N_{\rm CiC} \rangle)^k\rangle.
\end{equation}

For our analysis of the computational performance of these methods, and the tuning of hyperparameters introduced in Section~\ref{sec:pretab}, see Appendix~\ref{sec:appendix-computational-performance}

\startlongtable
\begin{deluxetable*}{c|c|c|c|c|c|c|c|c|c|c|c|c}
\tablecaption{Maximum-likelihood HOD parameters for each sample. For each set of best-fit parameters, the goodness of fit is given by the Akaike Information Criterion (AIC), the chi-squared ($\chi^2$), the degrees of freedom (DoF), the $p$ value corresponding to the probability of measuring $\geq \chi^2$ by chance, and the corresponding $z$ score measure of tension. The fits without CiC, and without assembly bias are included for comparison. \label{tab:hod-fits}}
\tablehead{
\colhead{Threshold} & \colhead{$\log M_{\rm min}$} & \colhead{$\sigma_{\log M}$} & \colhead{$\alpha$} & \colhead{$\log M_1$} & \colhead{$\log M_0$} & \colhead{$A_{\rm cen}$} & \colhead{$A_{\rm sat}$} & \colhead{AIC} & \colhead{$\chi^2$} & \colhead{DoF} & \colhead{$p$ value} & \colhead{Tension}
}
\startdata
-20.0 & 12.227 & 0.990 & 0.681 & 12.739 & 12.339 & 0.966 & -0.156 & -292.68 & 12.15 & 19 & 0.879 & $0.15\sigma$\\
(no CiC) & 12.114 & 0.884 & 0.858 & 12.946 & 12.430 & 0.540 & -0.795 & 49.66 & 10.20 & 13 & 0.678 & $0.42\sigma$\\
(no $A_{\rm bias}$) & 11.968 & 0.481 & 0.778 & 12.763 & 12.459 & &  & -284.95 & 23.88 & 19 & 0.201 & $1.28\sigma$\\[5pt]
-20.5 & 12.285 & 0.527 & 0.765 & 13.140 & 12.657 & 0.911 & -0.223 & -214.70 & 20.51 & 19 & 0.364 & $0.91\sigma$\\
(no CiC) & 12.923 & 1.387 & 0.566 & 12.935 & 12.930 & 0.164 & -0.317 & 52.74 & 7.70 & 13 & 0.863 & $0.17\sigma$\\
(no $A_{\rm bias}$) & 12.244 & 0.381 & 0.661 & 13.020 & 12.912 & &  & -208.36 & 30.85 & 19 & 0.042 & $2.03\sigma$\\[5pt]
-21.0 (low z) & 12.467 & 0.211 & 0.475 & 13.323 & 13.068 & 0.853 & 0.050 & -233.42 & 54.76 & 42 & 0.090 & $1.70\sigma$\\
(no CiC) & 12.411 & 0.063 & 0.819 & 13.618 & 12.643 & 0.885 & -0.249 & 58.89 & 3.88 & 13 & 0.992 & $0.01\sigma$\\
(no $A_{\rm bias}$) & 12.453 & 0.045 & 0.409 & 13.226 & 13.116 & &  & -234.72 & 57.46 & 42 & 0.056 & $1.91\sigma$\\[5pt]
-21.0 (high z) & 12.388 & 0.271 & 1.005 & 13.565 & 12.813 & 0.817 & -0.072 & -141.88 & 25.89 & 19 & 0.133 & $1.50\sigma$\\
(no CiC) & 12.415 & 0.398 & 0.758 & 13.475 & 12.836 & 0.890 & -0.549 & 57.89 & 17.13 & 13 & 0.193 & $1.30\sigma$\\
(no $A_{\rm bias}$) & 12.360 & 0.059 & 0.852 & 13.431 & 13.099 & &  & -136.33 & 35.43 & 19 & 0.012 & $2.50\sigma$\\[5pt]
\enddata
\end{deluxetable*}

\startlongtable
\begin{deluxetable*}{c|c|c|c|c|c|c|c}
\tablecaption{Confidence intervals of the HOD parameters from the 16th, 50th, and 84th percentiles of the marginalized posteriors. The confidence intervals without CiC constraints, and without assembly bias, are included for comparison. \label{tab:hod-constraints}}
\tablehead{
\colhead{Threshold} & \colhead{$\log M_{\rm min}$} & \colhead{$\sigma_{\log M}$} & \colhead{$\alpha$} & \colhead{$\log M_1$} & \colhead{$\log M_0$} & \colhead{$A_{\rm cen}$} & \colhead{$A_{\rm sat}$}
}
\startdata
-20.0 & $12.026_{-0.069}^{+0.087}$ & $0.587_{-0.136}^{+0.159}$ & $0.748_{-0.065}^{+0.059}$ & $12.833_{-0.094}^{+0.073}$ & $12.315_{-0.145}^{+0.163}$ & $0.848_{-0.210}^{+0.115}$ & $-0.028_{-0.226}^{+0.211}$\\
(no CiC) & $12.151_{-0.274}^{+1.047}$ & $0.845_{-0.635}^{+1.701}$ & $0.784_{-0.149}^{+0.125}$ & $12.833_{-0.287}^{+0.177}$ & $12.566_{-0.329}^{+0.156}$ & $0.613_{-0.556}^{+0.288}$ & $-0.260_{-0.423}^{+0.502}$\\
(no $A_{\rm bias}$) & $11.951_{-0.063}^{+0.080}$ & $0.454_{-0.164}^{+0.155}$ & $0.744_{-0.057}^{+0.063}$ & $12.759_{-0.084}^{+0.088}$ & $12.427_{-0.185}^{+0.127}$ & & \\[5pt]
-20.5 & $12.252_{-0.056}^{+0.074}$ & $0.471_{-0.122}^{+0.126}$ & $0.707_{-0.065}^{+0.065}$ & $13.102_{-0.104}^{+0.088}$ & $12.728_{-0.142}^{+0.121}$ & $0.862_{-0.205}^{+0.102}$ & $-0.113_{-0.222}^{+0.217}$\\
(no CiC) & $12.518_{-0.367}^{+1.300}$ & $0.916_{-0.715}^{+1.572}$ & $0.681_{-0.257}^{+0.182}$ & $13.094_{-0.500}^{+0.224}$ & $12.886_{-0.275}^{+0.152}$ & $0.462_{-0.771}^{+0.412}$ & $-0.072_{-0.576}^{+0.607}$\\
(no $A_{\rm bias}$) & $12.213_{-0.052}^{+0.074}$ & $0.389_{-0.172}^{+0.150}$ & $0.691_{-0.043}^{+0.055}$ & $13.017_{-0.065}^{+0.080}$ & $12.837_{-0.113}^{+0.067}$ & & \\[5pt]
-21.0 (low z) & $12.450_{-0.012}^{+0.015}$ & $0.083_{-0.057}^{+0.108}$ & $0.423_{-0.071}^{+0.108}$ & $13.292_{-0.100}^{+0.140}$ & $13.091_{-0.098}^{+0.046}$ & $0.229_{-0.758}^{+0.533}$ & $0.047_{-0.228}^{+0.154}$\\
(no CiC) & $12.464_{-0.038}^{+0.125}$ & $0.272_{-0.191}^{+0.293}$ & $0.719_{-0.232}^{+0.165}$ & $13.569_{-0.206}^{+0.112}$ & $12.871_{-0.253}^{+0.236}$ & $0.333_{-0.779}^{+0.501}$ & $-0.012_{-0.522}^{+0.562}$\\
(no $A_{\rm bias}$) & $12.455_{-0.011}^{+0.022}$ & $0.098_{-0.077}^{+0.160}$ & $0.414_{-0.097}^{+0.091}$ & $13.291_{-0.090}^{+0.119}$ & $13.080_{-0.088}^{+0.048}$ & & \\[5pt]
-21.0 (high z) & $12.365_{-0.027}^{+0.036}$ & $0.222_{-0.144}^{+0.126}$ & $0.895_{-0.090}^{+0.089}$ & $13.494_{-0.099}^{+0.095}$ & $12.944_{-0.173}^{+0.133}$ & $0.759_{-0.380}^{+0.185}$ & $-0.200_{-0.214}^{+0.200}$\\
(no CiC) & $12.356_{-0.024}^{+0.048}$ & $0.178_{-0.120}^{+0.175}$ & $0.959_{-0.118}^{+0.078}$ & $13.563_{-0.097}^{+0.052}$ & $12.597_{-0.154}^{+0.217}$ & $0.640_{-0.683}^{+0.276}$ & $-0.218_{-0.270}^{+0.252}$\\
(no $A_{\rm bias}$) & $12.366_{-0.025}^{+0.035}$ & $0.244_{-0.149}^{+0.118}$ & $0.929_{-0.064}^{+0.067}$ & $13.479_{-0.073}^{+0.078}$ & $12.964_{-0.143}^{+0.113}$ & & \\[5pt]
\enddata
\end{deluxetable*}

\section{MCMC Inference}
\label{sec:constraints}

\begin{figure*}
    \centering
    \includegraphics[width=\textwidth]{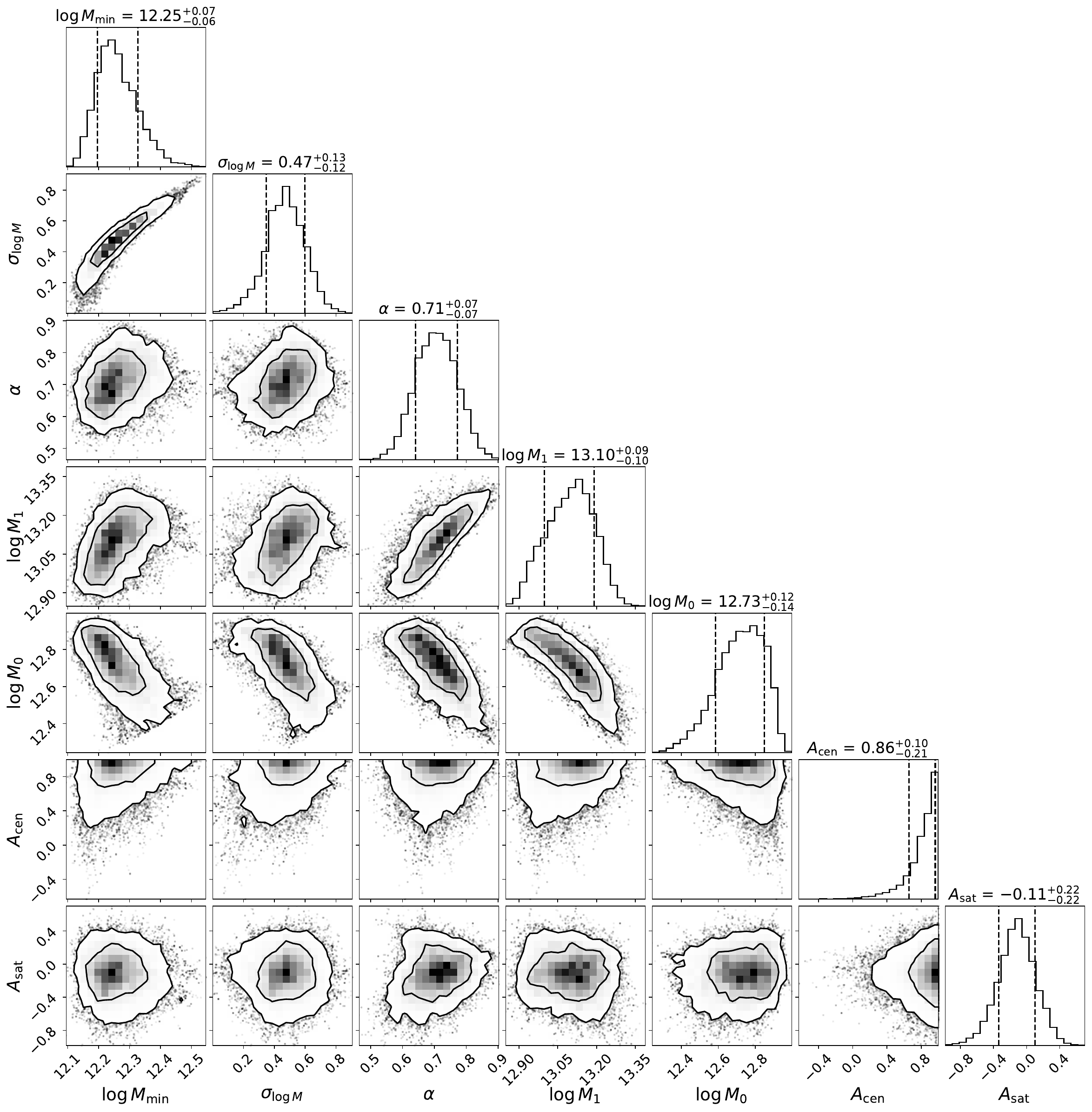}
    \caption{Posterior distribution of the HOD parameters of the -20.5 threshold sample from MCMC sampling. The 68\% and 95\% confidence regions are displayed by contour lines for each two-dimensional projection, and the 68\% confidence intervals are marked with dashed vertical lines for each one-dimensional projection.}
    \label{fig:mcmc-posterior}
\end{figure*}

\begin{figure}
    \centering
    \includegraphics[width=0.47\textwidth]{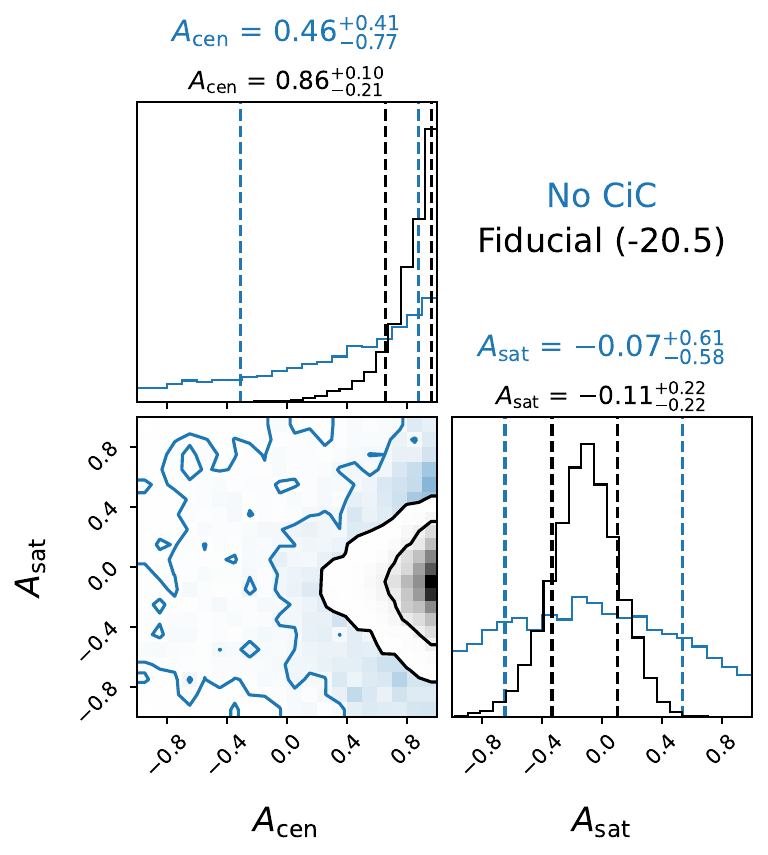}
    \caption{Posterior distribution of the assembly bias parameters of the -20.5 threshold sample from MCMC sampling. Overplotted in blue is the result we obtain without including any constraints from CiC, yielding very little information about assembly bias.}
    \label{fig:assembias-vs-kmax0}
\end{figure}

We use Markov-chain Monte Carlo (MCMC) to constrain the HOD model using each galaxy sample. We make use of the \texttt{emcee} \citep{Foreman-Mackey:2013} implementation, in which several walkers simultaneously sample a likelihood function throughout parameter space, and occasionally trade locations to construct MCMC chains. Ignoring the normalization constant, the log-likelihood is given by

\begin{equation}
    \ln \mathcal{L} = -\frac{1}{2} (\vec{x}_{\rm model} - \vec{x}_{\rm data})^\mathsf{T} \Sigma^{\dagger} (\vec{x}_{\rm model} - \vec{x}_{\rm data})
\end{equation}
where $\Sigma$ is the covariance matrix from Equation~\ref{eq:cov-jackknife} and $\Sigma^{\dagger}$ is its Moore-Penrose pseudo-inverse \citep{Penrose:1955}, which
is the simplest way to invert a singular matrix to calculate sensible, finite likelihood values, by performing a dimensionality reduction.
A singular covariance matrix arises when there are at least as many summary statistics as the number of jackknife realizations. We use the implementation available in the \texttt{logpdf} method of the \texttt{multivariate\_normal} class from SciPy \citep{Virtanen:2020}.

In addition, we rescale the summary statistics such that their covariance matrix has a diagonal of ones. Mathematically, this has no effect and is equivalent to an arbitrary change of units. However, this circumvents machine precision errors where the pseudo-inverse will delete the constraints of summary statistics with low orders of magnitude, like number density.

We initialize our MCMC chains around the best-fit parameters of the corresponding magnitude threshold sample from \citet{Wang:2022}, with very slight variation between the MCMC walkers. We let these chains run for 60,000 trial points (3,000 iterations $\times$ 20 walkers), and conservatively remove a burn-in of 2,000 trial points to calculate our posteriors displayed in Figures~\ref{fig:mcmc-posterior}, \ref{fig:assembias-vs-kmax0}, and~\ref{fig:assembias-vs-htcic}, as well as the maximum-likelihood points and confidence regions reported in Tables~\ref{tab:hod-fits} and~\ref{tab:hod-constraints}, respectively. Our relatively small number of trial points is acceptable thanks to our deterministic likelihood evaluations and our prior on $\log M_0$ that confines the MCMC to a stable region of parameter space. The autocorrelation lengths of our chains ended up ranging from 100-300. This is about a factor of two shorter than the autocorrelation lengths we obtain using Monte Carlo CiC evaluations, and possibly an order of magnitude shorter than the result from Monte Carlo evaluations of both \wprp\ and CiC.

To quantify how well our maximum-likelihood models agree with the data, we calculate $\chi^2$ along with the probability of measuring data with at least this value of $\chi^2$ by chance using the chi-squared cumulative distribution function. In Table~\ref{tab:hod-fits}, we report this probability and translate it into the $z$-score of a Gaussian to quantify the ``number of sigmas'' of tension that exists between our model and data.

\begin{figure*}[ht!]
    \centering
    \includegraphics[width=0.95\textwidth]{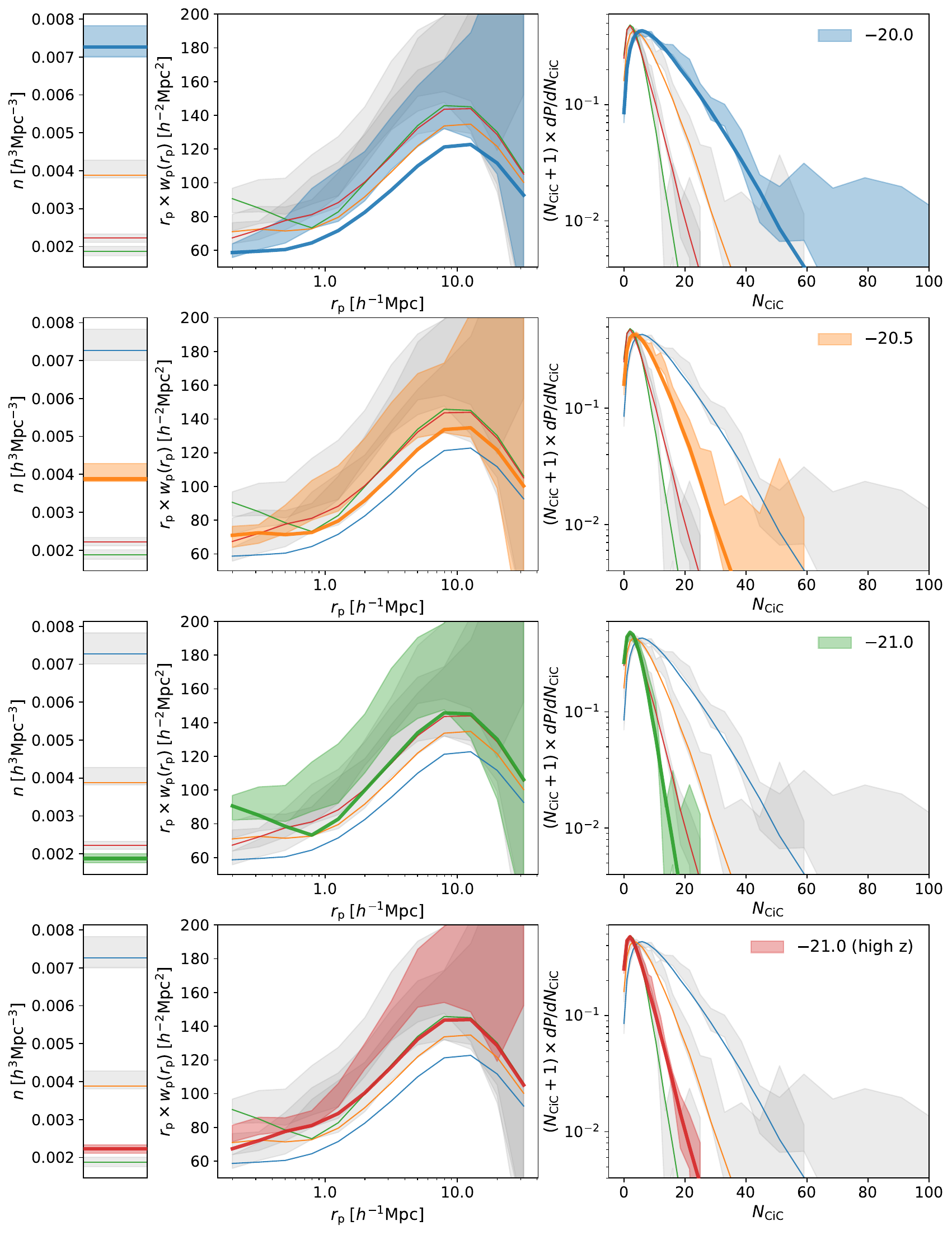}
    \caption{DESI measurements and our maximum-likelihood predictions of number density (left panels), the projected correlation function (center panels), and the CiC distribution (right panels). The 1$\sigma$ confidence intervals from the measurements of a given quantity are represented by shaded regions of the color corresponding to the sample, while the solid lines, following the same color scheme, represent our model's maximum-likelihood predictions. The parameters corresponding to these best-fit predictions are reported in Table~\ref{tab:hod-fits}.}
    \label{fig:summary-stats}
\end{figure*}

\section{Results and Discussion}

\begin{figure}
    \centering
    \includegraphics[width=0.47\textwidth]{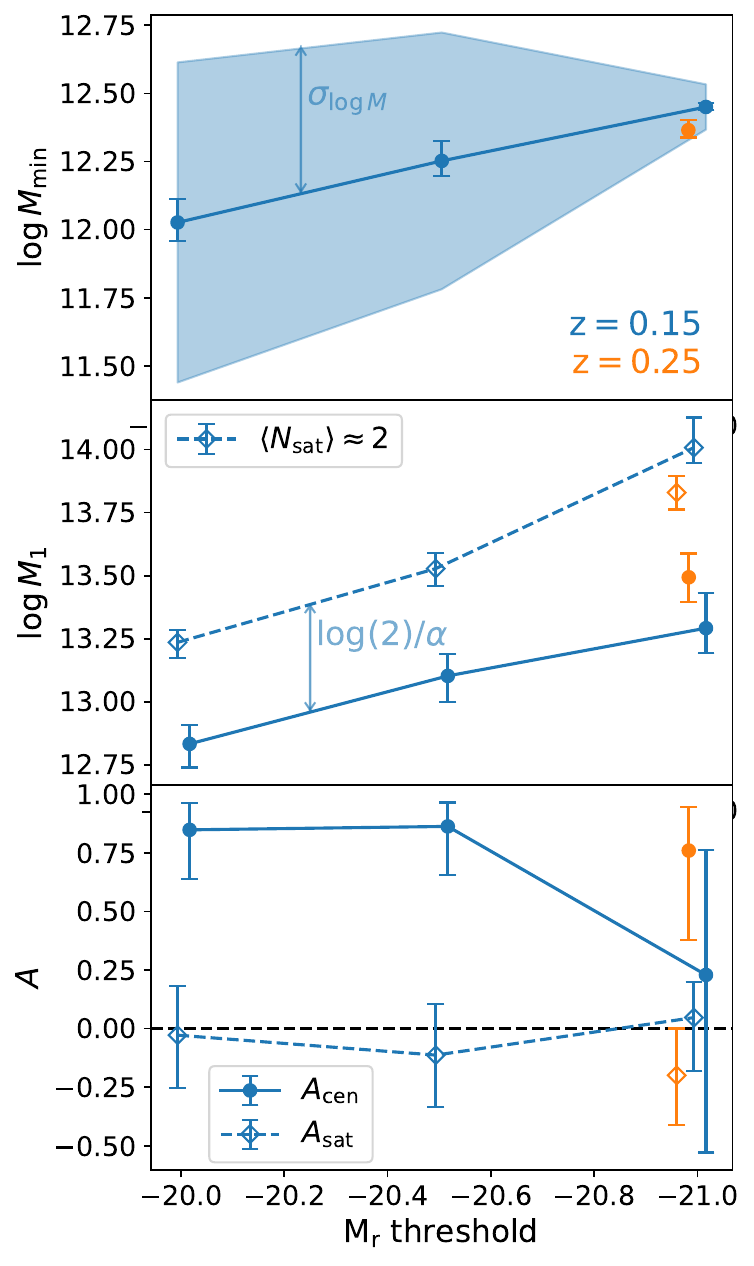}
    \caption{Variation of HOD parameters with luminosity and redshift. Median values of the one-dimensional marginalized posteriors for the characteristic masses, $\log M_{\rm min}$ (top panel) and $\log M_1$ (middle panel) are plotted, as well as the assembly bias parameters $A_{\rm cen}$ and $A_{\rm sat}$ (bottom panel). The capped error bars on these points span the 16th to the 84th percentile of the posterior for a given parameter. Median values derived from our posteriors of other HOD parameters $\sigma_{\log M}$ (top panel) and $\alpha$ (middle panel) are labeled; $\sigma_{\log M}$ characterizes the spread in the $M_r$-$M_{\rm halo}$ relation, and $\log(2)/\alpha$ characterizes the log-difference between the halo masses corresponding to $\langle N_{\rm sat} \rangle \approx 1$ and $\langle N_{\rm sat} \rangle \approx 2$. We apply small x-offsets to distinguish the points easily, but all $M_r$ thresholds are exactly $-20.0$, $-20.5$, or $-21.0$.}
    \label{fig:hod-evolution}
\end{figure}

The measurements from the DESI One-Percent Survey already produce reasonably tight constraints on the HOD. For each of the four threshold samples defined in Table~\ref{tab:sample-cuts}, the corresponding best-fit HOD parameters are given in Table~\ref{tab:hod-fits}, and $1\sigma$ confidence intervals are given in Table~\ref{tab:hod-constraints}.
We have also summarized these constraints as a function of $M_r$ threshold and redshift into easier-to-digest plots in Figure~\ref{fig:hod-evolution}. In this figure, we show that as luminosity increases from $M_r$ of $-20.0$ to $-21.0$, the characteristic halo mass for central galaxies gradually increases from roughly $10^{12.0}$ to $10^{12.4}~M_{\odot}$. We find a similar increasing trend for the characteristic halo masses containing one (and two) satellite galaxies for each sample; the inferred slope $\alpha$ of the $\langle N_{\rm sat} \rangle(M_{\rm halo})$ relation does not evolve significantly compared to the shown error bars. Finally, we show the parameters which trace assembly bias; these are very significantly greater than zero for centrals in the lower two magnitude threshold samples, while satellite assembly bias is consistent with zero throughout. With only one $z=0.25$ sample, we find no significant signals of redshift evolution.

Given the current relatively small sample sizes, the tightness of our constraints can be attributed to the power of combining information from $w_{\rm p}$ and CiC. We find a $3\sigma$ detection of assembly bias for central galaxies in the two lower luminosity bins. More precisely, the strength of the evidence for central assembly bias in each sample is as follows:
\begin{itemize}
    \item For our $-20.0$ and $-20.5$ samples, the posterior probability that $A_{\rm cen} > 0$ is 0.9987 and 0.995, respectively.  Without CiC constraints, these probabilities are only 0.860 and 0.737.
    \item Positive assembly bias at $M_r < -21.0$ is favored significantly only in the $z\sim0.25$ sample. For it, we find a posterior  probability for $A_{\rm cen} > 0$ of 0.948 (or 0.828 without CiC constraints).
    \item Due to large uncertainties, we find very poor constraints on assembly bias at $M_r < -21.0$  in our $z\sim0.15$ sample whether or not we include CiC in the sample.
    \item In general, CiC appears to add a substantial increase in constraining power for all HOD parameters, as seen in Figure~\ref{fig:feature-importance} and in detail in Table~\ref{tab:hod-constraints}.
\end{itemize}

The constraints we find on assembly bias are consistent with the findings from studies based on SDSS data. Despite the smaller sample size currently available from DESI, our \wprp\ + CiC analysis produces much stronger constraints than characterizing SDSS clustering with \wprp\ alone \citep[e.g.,][]{Zentner:2019, Vakili:Hahn:2019}. In fact, we achieve similar constraining power to \citet{Wang:2022}, even though we use the same set of summary statistics. This may imply that the assembly bias signal is less ambiguous in the slightly different samples probed by BGS. This could also be thanks to the high targeting completeness and therefore purity of the DESI One-Percent Survey, which allows us to avoid assigning redshifts to untargeted galaxies based upon the nearest neighbors in the sky. Finally, if the HOD model is not sufficiently flexible (a good possibility given our imperfect fits), our results will be prior dominated, which can affect the inference in unpredictable ways.

While the HOD model can consistently produce good fits to $w_{\rm p}$ and $n$ simultaneously (possibly to the point of overfitting), incorporating CiC measurements results in mismatches between the model and data in some cases. Although introducing assembly bias parameters has slightly reduced this tension, the $M_r < -21.0$ sample at $z\sim0.15$ still exhibits a tension of nearly 2$\sigma$ between our models and the data. This tension is reported in Table~\ref{tab:hod-fits} and is readily apparent in Figure~\ref{fig:summary-stats} (though one must use caution when assessing the mismatch by eye since the summary statistics can be strongly covariant).

Significant tension in only one of our four samples by no means rules out the HOD model used, but it should incentivize us to consider what else the model might be missing.  In the coming years, the size of the DESI sample will grow by a factor of 100 compared to what was used here, so we can expect that the constraints will tighten significantly and tensions may grow. Our model is not sufficiently flexible to fit early data samples well; therefore, it is plausible that these models could be ruled out convincingly with the full dataset. Future studies should explore additional ways to make the HOD more flexible such that they can produce better fits to the DESI data; we describe a few plausible extensions here, but by no means exhaust the possibilities.
\label{sec:results}

As one example, the HOD we have used in this work assumes that the stellar-to-halo-mass relation has a log-normal scatter, but the UniverseMachine simulations \citep{Behroozi:2019} exhibit a slight skew to this scatter in several tested samples. In principle, it is simple to test the addition of one more parameter to allow a skew-log-normal scatter. This would give HOD models the ability to capture some of the flexibility built into more sophisticated models.

Another modification that may be justified is to relax the assumed isotropic NFW distribution of satellite galaxies. This is a common assumption, yet it has long been known that the distribution of subhalos is anisotropic, due to the preferential accretion of mergers along filaments \citep{Zentner:2005}. Additionally, recent studies have found a significant difference in the radial profile of the halo mass associated with subhalos from NFW \citep{Fielder:2020, Mezini:2023}. Such modifications would be more complex but will be particularly important as small-scale clustering measurements improve since they are sensitive to the spatial distribution of satellites. However, it is possible that the satellite profile is degenerate with assembly bias for CiC. Therefore, any modifications to the satellite profile should be validated against high-resolution subhalo profiles.

Additionally, we have only tested for assembly bias tied to halo concentration and have ignored other occupation correlations that may be based upon halo spin, age, or environment \citep{Contreras:2021, Sato-Polito:2019, Yuan:2021}. Another possibility is that the occupation of satellites is correlated with the occupation of the central in the same halo due to galactic conformity \citep{Berti:2017, Kauffmann:2013}. All of these scenarios would likely produce similar statistical imprints. However, a primary question to investigate is whether these alternate assumptions lead to a biased inference of HOD parameters such as characteristic halo masses and assembly bias. If so, all of our results could be overly confident\footnote{i.e., Zentner Points\texttrademark}.

While CiC plays a crucial role in the HOD constraints obtained via our analysis, it is also our computational bottleneck. However, we have significantly sped up this process with \texttt{galtab}, particularly by removing the stochasticity of likelihood evaluations, which greatly improves the MCMC convergence rate. Using a stochastic estimator, convergence is especially problematic for the lowest-number-density, brightest-threshold samples, which exhibit order-of-magnitude increases in the acceptance rates of their MCMC chains.

Depending on the computing resources available and the dimensionality of the analysis, \texttt{galtab} may provide even more drastic speedups. Due to the implementation in JAX, the expensive steps are automatically executed on a GPU when available. Additionally, our framework allows the predictions to be differentiable with respect to HOD parameters (assuming the occupation model is compatible with JAX arrays, for which those available in \texttt{halotools} require slight modifications). In principle, this allows for the use of alternative MCMC methods with improved scalability to high-dimensional or strongly covariant posterior estimation, such as Hamiltonian Monte Carlo \citep{Neal:2011}.

Our development of the \texttt{galtab} package provides a useful tool for further analyses of the galaxy-halo connection that may require differentiable predictions. By combining these new tools with upcoming enlarged samples from DESI, we anticipate that coming studies will soon shift focus from mere detections of assembly bias to studying its implications for galaxy formation in much finer detail.

\appendix

\section{SHAP Feature Importance Calculations}
\label{sec:appendix-shap}

As briefly discussed in Section~\ref{sec:summary-stats} and plotted in Figure~\ref{fig:feature-importance}, we have roughly quantified the importance of each summary statistic in inferring the parameters of the HOD model by testing how influential each quantity is for predictions based on machine learning. We performed this test using an artificial dataset based upon uniformly sampling 1000 sets of all seven HOD parameters (see Section~\ref{sec:hod}) by Latin Hypercube Sampling over the projected one-dimensional $1\sigma$ confidence interval of the fiducial fits for the $M_r < -20.5$ threshold sample of \citet{Wang:2022}.

For each of the 1000 sets of HOD parameters, we predicted the values of all the summary statistics listed in Section~\ref{sec:summary-stats} using the methods described in Section~\ref{sec:analytic-prediction}. For each evaluation, we incorporated artificial observational uncertainty from a draw of our $M_r < -20.5$ covariance matrix. We then trained a scikit-learn \citep{Pedregosa:2012} random forest regression model to perform the inverse mapping (i.e., predict HOD parameters from the values of the summary statistics).

We then calculate SHAP feature importance values for each feature in the random forest. SHAP values are explained in detail in \citet{Lundberg:2017}. In brief, they attempt to map feature values to their linear ``impact'' on model predictions. For example, a large positive SHAP value is assigned to a feature value that produced a large increase in the model prediction, while a large negative SHAP value is assigned to a feature value that produced a large decrease in the model prediction. This allows us to analyze and distinguish the effects of positive or negative changes in each feature on the model predictions that result.

\begin{figure*}[ht!]
    \centering
    \includegraphics[width=\textwidth]{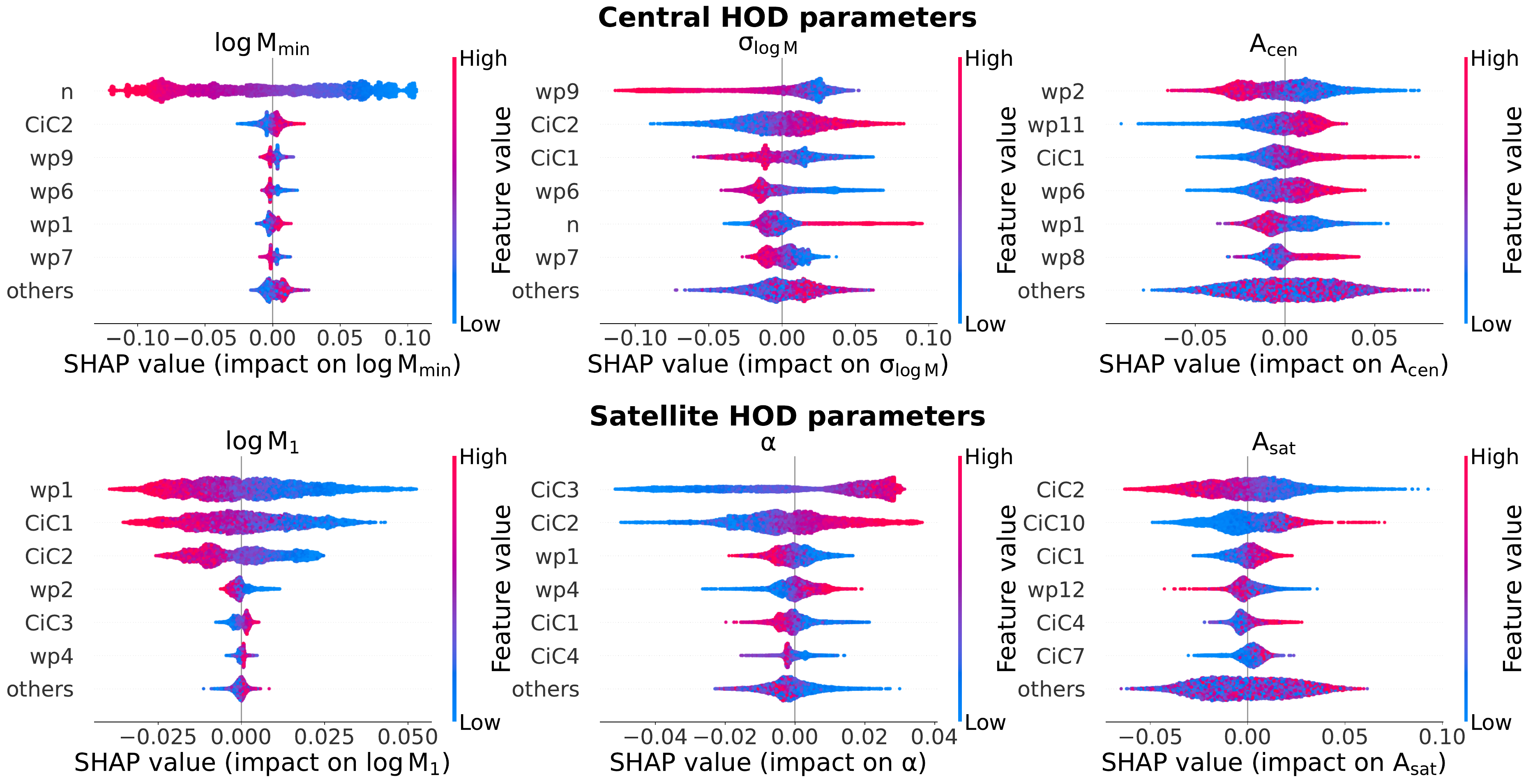}
    \caption{The impact of each of our summary statistics on HOD inference, based upon SHAP feature importances. The upper panel of each sub-figure shows beeswarms of the SHAP values for each feature's impact on predicting the given HOD parameter. Each panel shows the six most important quantities in order of importance, and the panels are organized in the same way those in Figure~\ref{fig:feature-importance}. See Figure~\ref{fig:feature-importance} for a more condensed version of this information which focuses on the mean absolute SHAP value as an importance metric.}
    \label{fig:shap-beeswarm}
\end{figure*}

We show the complete beeswarm distribution of SHAP values corresponding to the variation of each HOD parameter in Figure~\ref{fig:shap-beeswarm}. We calculate the importance values shown in Figure~\ref{fig:feature-importance} by taking the mean absolute values of these distributions. Features with large importance values correspond to quantities most useful for predicting a given HOD parameter. We roughly estimate importance uncertainties from the standard error of the mean of each set of 1000 absolute SHAP values. However, note that this estimate does not account for sample variance in the simulation volume nor for systematics that might arise from the SHAP formulation.

This experiment demonstrates how informative CiC is to our results. In particular, the vast majority of information content comes from the first three moments alone. The only exception to this statement appears to be the tenth CiC moment in predicting $A_{\rm sat}$, but further testing has shown that this is not a real artifact. In fact, we find an artificial boost in $A_{\rm sat}$ importance on the highest CiC moment, no matter how high we go to. CiC appears to be especially crucial for informing the satellite HOD parameters, likely due to the small scale of our cylinders, and the first few moments have significant importance across every single parameter.

\section{Computational Performance}
\label{sec:appendix-computational-performance}

\begin{figure}[ht!]
    \centering
    \includegraphics[width=0.47\textwidth]{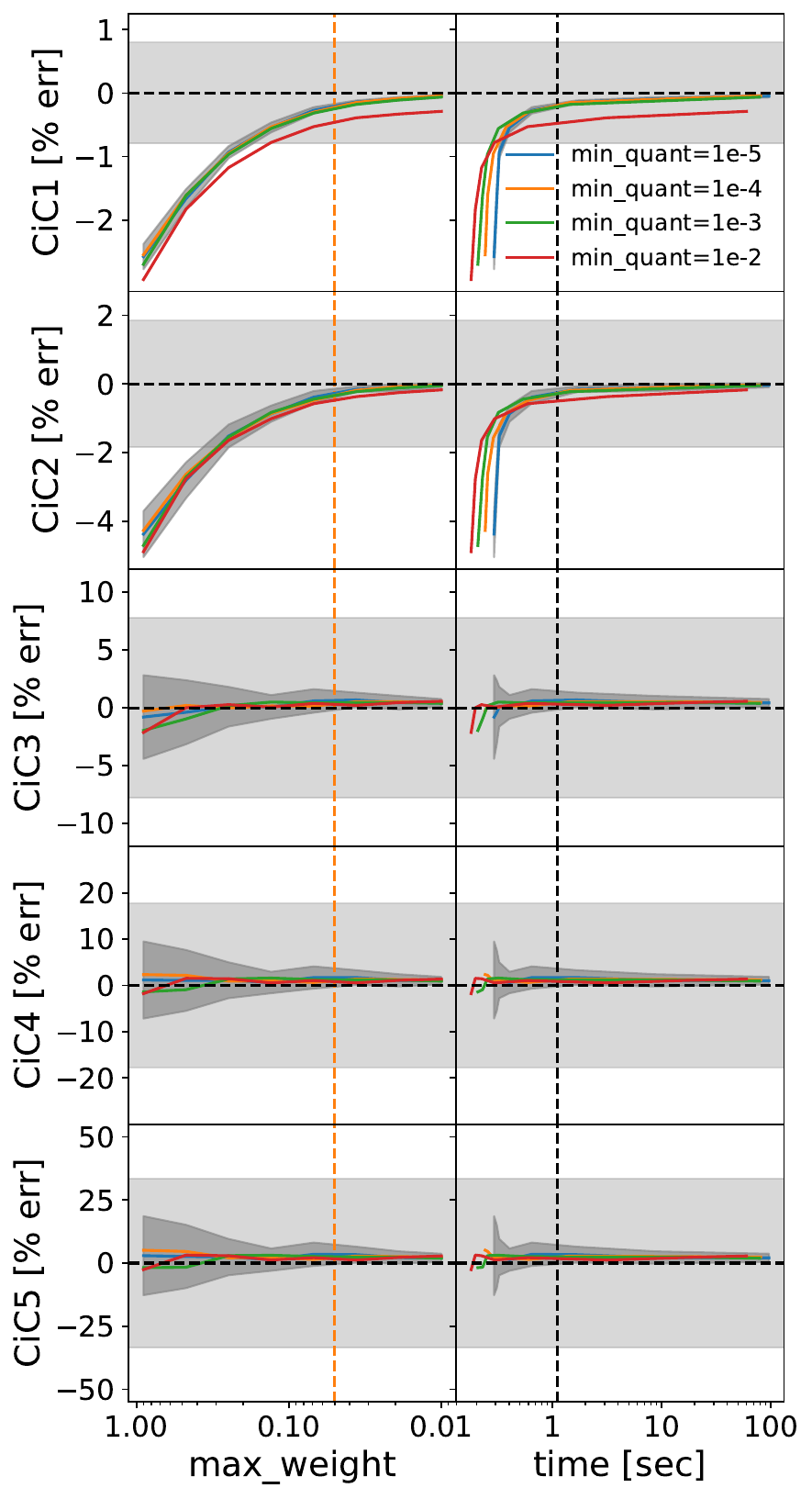}
    \caption{Hyperparameter tuning of \texttt{galtab} to achieve sufficient accuracy of CiC moments. The left panels show the tuning of the $W_{\rm max}$ parameter, which is translated to a CPU runtime in the panels on the right side of the figure, with lower values of $W_{\rm max}$ requiring longer times, but achieving higher accuracy. Line colors correspond to the denoted value of $Q_{\rm min}$, the dark grey bands correspond to a standard deviation due to tabulation stochasticity, horizontal dashed lines correspond to truth values from \texttt{halotools}, and the light grey band corresponds to a \texttt{halotools} standard deviation. The vertical dashed line in the left panels corresponds to our chosen value of $W_{\rm max}=0.05$, which intentionally yields a similar runtime to \texttt{halotools}: approximately one CPU-second, as specified by the vertical dashed line in the right panels.
    }
    \label{fig:accuracy-runtime}
\end{figure}

In Section~\ref{sec:pretab} and Figure~\ref{fig:accuracy-runtime}, we have described our hyperparameter tuning of $W_{\rm max}$ and $Q_{\rm min}$ to balance runtime and accuracy. These parameters control the number of placeholders, $N$, as well as the average number of placeholders per cylinder, $C$. To store all pretabulated indices, the memory usage of \texttt{galtab} scales with $\mathcal{O}(NC)$.

There are also additional runtime considerations specific to each prediction mode. For the Monte Carlo mode, the runtime scales with the number of effective Monte Carlo realizations, $n_{\rm MC}$, so the time complexity is $\mathcal{O}(n_{\rm MC} N C)$. For the analytic mode, the runtime scales with the highest calculated moment, $k_{\rm max}$, so the time complexity is $\mathcal{O}(k_{\rm max} N C)$.

By far, the most computationally expensive step of our procedure is the summation of occupations (or cumulants, for the analytic mode; see Equation~\ref{eq:cumulant-sum}) of placeholders per cylinder. To fully optimize this calculation, we employ just-in-time (JIT) compilation via the JAX library \citep{JaxGitHub:2018}. This also automatically ports the computation to the GPU, if available, which can speed up the predictions by at least an order of magnitude faster than the times reported in Figure~\ref{fig:accuracy-runtime}.

\begin{figure}
    \centering
    \includegraphics[width=0.47\textwidth]{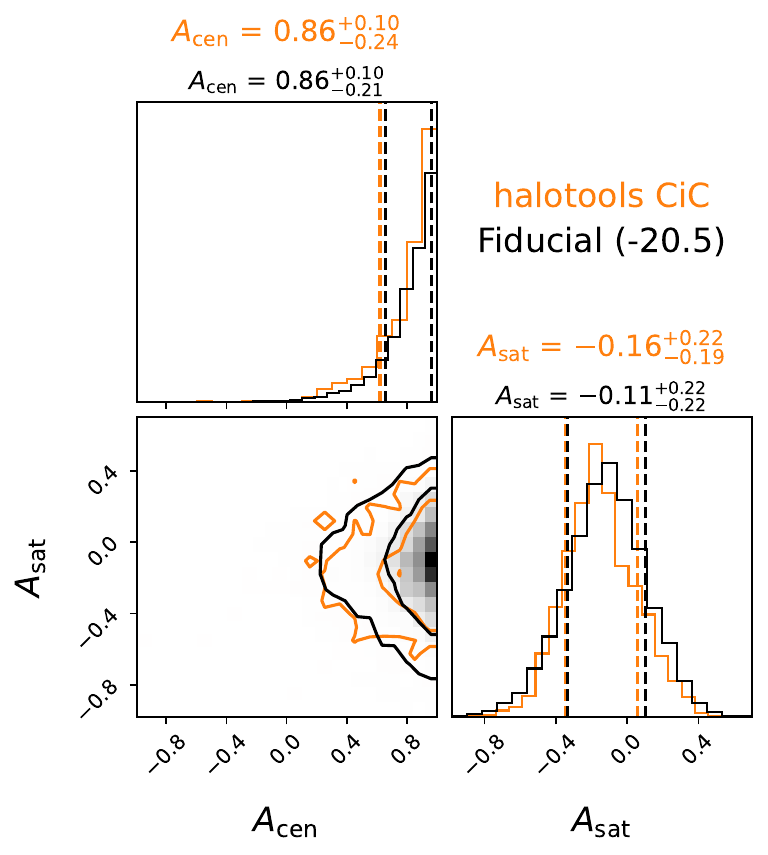}
    \caption{Posterior distribution of the assembly bias parameters of the -20.5 threshold sample from MCMC sampling. Overplotted in orange is the result we obtain from calculating CiC from \texttt{halotools} instead of \texttt{galtab} for the same number of MCMC iterations. Due to the stochasticity of the \texttt{halotools} predictions, its acceptance rate was three times lower in this case, causing much slower posterior convergence.}
    \label{fig:assembias-vs-htcic}
\end{figure}

The primary advantage of \texttt{galtab} over the Monte Carlo prediction methods available in \texttt{halotools} is that its predictions are deterministic. After performing the tabulation, the same inputs will always yield the same outputs (and there is not much scatter between different tabulation realizations, as seen in Figure~\ref{fig:accuracy-runtime}). Deterministic likelihood function calls yield much more efficient MCMC convergence, thanks to higher acceptance rates, and lower autocorrelation lengths. We have tested the difference in posterior inference between \texttt{galtab} and \texttt{halotools} in Figure~\ref{fig:assembias-vs-htcic}. Each of these trials performed the same number of MCMC iterations (60,000 trial points), and took essentially the same amount of time, but \texttt{galtab} produces much smoother contour lines, which are indicative of a more well-converged posterior.

\section*{Data Availability}

Our figure data, including summary statistics, covariance matrices, and MCMC results for each of our HOD samples are available to download at \url{https://doi.org/10.5281/zenodo.8206461}.

\acknowledgements

This material is based upon work supported by the U.S. Department of Energy (DOE), Office of Science, Office of High-Energy Physics, under Contract No. DE–AC02–05CH11231, and by the National Energy Research Scientific Computing Center, a DOE Office of Science User Facility under the same contract. Additional support for DESI was provided by the U.S. National Science Foundation (NSF), Division of Astronomical Sciences under Contract No. AST-0950945 to the NSF’s National Optical-Infrared Astronomy Research Laboratory; the Science and Technologies Facilities Council of the United Kingdom; the Gordon and Betty Moore Foundation; the Heising-Simons Foundation; the French Alternative Energies and Atomic Energy Commission (CEA); the National Council of Science and Technology of Mexico (CONACYT); the Ministry of Science and Innovation of Spain (MICINN), and by the DESI Member Institutions: \url{https://www.desi.lbl.gov/collaborating-institutions}. Any opinions, findings, and conclusions or recommendations expressed in this material are those of the author(s) and do not necessarily reflect the views of the U. S. National Science Foundation, the U. S. Department of Energy, or any of the listed funding agencies.

The authors are honored to be permitted to conduct scientific research on Iolkam Du’ag (Kitt Peak), a mountain with particular significance to the Tohono O’odham Nation.

The efforts of J. A. Newman were supported by grant DE-SC0007914 from the U.S. Department of Energy Office of Science, Office of High Energy Physics.

This research has made extensive use of the arXiv and NASA's Astrophysics Data System.
This research has made use of adstex (\url{https://github.com/yymao/adstex}).

\software{
Halotools \citep{Hearin:2016},
JAX \citep{JaxGitHub:2018},
Corrfunc \citep{Sinha:2020},
emcee \citep{Foreman-Mackey:2013},
corner.py \citep{Foreman-Mackey:2016},
scikit-learn \citep{Pedregosa:2012},
SciPy \citep{Virtanen:2020},
matplotlib \citep{Hunter:2007},
Astropy \citep{Astropy:2022, Astropy:2013},
NumPy \citep{vanderWalt:2011},
}

\vspace{1000pt}

\bibliography{bibtex}

\begin{thebibliography}{}
\expandafter\ifx\csname natexlab\endcsname\relax\def\natexlab#1{#1}\fi
\providecommand{\url}[1]{\href{#1}{#1}}
\providecommand{\dodoi}[1]{doi:~\href{http://doi.org/#1}{\nolinkurl{#1}}}
\providecommand{\doeprint}[1]{\href{http://ascl.net/#1}{\nolinkurl{http://ascl.net/#1}}}
\providecommand{\doarXiv}[1]{\href{https://arxiv.org/abs/#1}{\nolinkurl{https://arxiv.org/abs/#1}}}

\bibitem[{{Abazajian} {et~al.}(2009){Abazajian}, {Adelman-McCarthy}, {Ag{\"u}eros}, {Allam}, {Allende Prieto}, {An}, {Anderson}, {Anderson}, {Annis}, {Bahcall}, \& et~al.}]{SDSS:2009}
{Abazajian}, K.~N., {Adelman-McCarthy}, J.~K., {Ag{\"u}eros}, M.~A., {et~al.} 2009, \apjs, 182, 543, \dodoi{10.1088/0067-0049/182/2/543}

\bibitem[{{Abbott} {et~al.}(2018){Abbott}, {Abdalla}, {Alarcon}, {Aleksi{\'c}}, {Allam}, {Allen}, {Amara}, {Annis}, {Asorey}, {Avila}, \& et~al.}]{Abbott:2018}
{Abbott}, T.~M.~C., {Abdalla}, F.~B., {Alarcon}, A., {et~al.} 2018, \prd, 98, 043526, \dodoi{10.1103/PhysRevD.98.043526}

\bibitem[{{Adelberger} {et~al.}(1998){Adelberger}, {Steidel}, {Giavalisco}, {Dickinson}, {Pettini}, \& {Kellogg}}]{Adelberger:1998}
{Adelberger}, K.~L., {Steidel}, C.~C., {Giavalisco}, M., {et~al.} 1998, \apj, 505, 18, \dodoi{10.1086/306162}

\bibitem[{{Anderson} {et~al.}(2012){Anderson}, {Aubourg}, {Bailey}, {Bizyaev}, {Blanton}, {Bolton}, {Brinkmann}, {Brownstein}, {Burden}, {Cuesta}, {da Costa}, {Dawson}, {de Putter}, {Eisenstein}, {Gunn}, {Guo}, {Hamilton}, {Harding}, {Ho}, {Honscheid}, {Kazin}, {Kirkby}, {Kneib}, {Labatie}, {Loomis}, {Lupton}, {Malanushenko}, {Malanushenko}, {Mandelbaum}, {Manera}, {Maraston}, {McBride}, {Mehta}, {Mena}, {Montesano}, {Muna}, {Nichol}, {Nuza}, {Olmstead}, {Oravetz}, {Padmanabhan}, {Palanque-Delabrouille}, {Pan}, {Parejko}, {P{\^a}ris}, {Percival}, {Petitjean}, {Prada}, {Reid}, {Roe}, {Ross}, {Ross}, {Samushia}, {S{\'a}nchez}, {Schlegel}, {Schneider}, {Sc{\'o}ccola}, {Seo}, {Sheldon}, {Simmons}, {Skibba}, {Strauss}, {Swanson}, {Thomas}, {Tinker}, {Tojeiro}, {Maga{\~n}a}, {Verde}, {Wagner}, {Wake}, {Weaver}, {Weinberg}, {White}, {Xu}, {Y{\`e}che}, {Zehavi}, \& {Zhao}}]{Anderson:2012}
{Anderson}, L., {Aubourg}, E., {Bailey}, S., {et~al.} 2012, \mnras, 427, 3435, \dodoi{10.1111/j.1365-2966.2012.22066.x}

\bibitem[{{Astropy Collaboration} {et~al.}(2013){Astropy Collaboration}, {Robitaille}, {Tollerud}, {Greenfield}, {Droettboom}, {Bray}, {Aldcroft}, {Davis}, {Ginsburg}, {Price-Whelan}, {Kerzendorf}, {Conley}, {Crighton}, {Barbary}, {Muna}, {Ferguson}, {Grollier}, {Parikh}, {Nair}, {Unther}, {Deil}, {Woillez}, {Conseil}, {Kramer}, {Turner}, {Singer}, {Fox}, {Weaver}, {Zabalza}, {Edwards}, {Azalee Bostroem}, {Burke}, {Casey}, {Crawford}, {Dencheva}, {Ely}, {Jenness}, {Labrie}, {Lim}, {Pierfederici}, {Pontzen}, {Ptak}, {Refsdal}, {Servillat}, \& {Streicher}}]{Astropy:2013}
{Astropy Collaboration}, {Robitaille}, T.~P., {Tollerud}, E.~J., {et~al.} 2013, \aap, 558, A33, \dodoi{10.1051/0004-6361/201322068}

\bibitem[{{Astropy Collaboration} {et~al.}(2022){Astropy Collaboration}, {Price-Whelan}, {Lim}, {Earl}, {Starkman}, {Bradley}, {Shupe}, {Patil}, {Corrales}, {Brasseur}, {N{\"o}the}, {Donath}, {Tollerud}, {Morris}, {Ginsburg}, {Vaher}, {Weaver}, {Tocknell}, {Jamieson}, {van Kerkwijk}, {Robitaille}, {Merry}, {Bachetti}, {G{\"u}nther}, {Aldcroft}, {Alvarado-Montes}, {Archibald}, {B{\'o}di}, {Bapat}, {Barentsen}, {Baz{\'a}n}, {Biswas}, {Boquien}, {Burke}, {Cara}, {Cara}, {Conroy}, {Conseil}, {Craig}, {Cross}, {Cruz}, {D'Eugenio}, {Dencheva}, {Devillepoix}, {Dietrich}, {Eigenbrot}, {Erben}, {Ferreira}, {Foreman-Mackey}, {Fox}, {Freij}, {Garg}, {Geda}, {Glattly}, {Gondhalekar}, {Gordon}, {Grant}, {Greenfield}, {Groener}, {Guest}, {Gurovich}, {Handberg}, {Hart}, {Hatfield-Dodds}, {Homeier}, {Hosseinzadeh}, {Jenness}, {Jones}, {Joseph}, {Kalmbach}, {Karamehmetoglu}, {Ka{\l}uszy{\'n}ski}, {Kelley}, {Kern}, {Kerzendorf}, {Koch}, {Kulumani}, {Lee}, {Ly}, {Ma}, {MacBride}, {Maljaars}, {Muna}, {Murphy}, {Norman},
  {O'Steen}, {Oman}, {Pacifici}, {Pascual}, {Pascual-Granado}, {Patil}, {Perren}, {Pickering}, {Rastogi}, {Roulston}, {Ryan}, {Rykoff}, {Sabater}, {Sakurikar}, {Salgado}, {Sanghi}, {Saunders}, {Savchenko}, {Schwardt}, {Seifert-Eckert}, {Shih}, {Jain}, {Shukla}, {Sick}, {Simpson}, {Singanamalla}, {Singer}, {Singhal}, {Sinha}, {Sip{\H{o}}cz}, {Spitler}, {Stansby}, {Streicher}, {{\v{S}}umak}, {Swinbank}, {Taranu}, {Tewary}, {Tremblay}, {de Val-Borro}, {Van Kooten}, {Vasovi{\'c}}, {Verma}, {de Miranda Cardoso}, {Williams}, {Wilson}, {Winkel}, {Wood-Vasey}, {Xue}, {Yoachim}, {Zhang}, {Zonca}, \& {Astropy Project Contributors}}]{Astropy:2022}
{Astropy Collaboration}, {Price-Whelan}, A.~M., {Lim}, P.~L., {et~al.} 2022, \apj, 935, 167, \dodoi{10.3847/1538-4357/ac7c74}

\bibitem[{{Behroozi} {et~al.}(2019){Behroozi}, {Wechsler}, {Hearin}, \& {Conroy}}]{Behroozi:2019}
{Behroozi}, P., {Wechsler}, R.~H., {Hearin}, A.~P., \& {Conroy}, C. 2019, \mnras, 488, 3143, \dodoi{10.1093/mnras/stz1182}

\bibitem[{{Behroozi} {et~al.}(2013){Behroozi}, {Wechsler}, \& {Wu}}]{Behroozi:2013}
{Behroozi}, P.~S., {Wechsler}, R.~H., \& {Wu}, H.-Y. 2013, \apj, 762, 109, \dodoi{10.1088/0004-637X/762/2/109}

\bibitem[{{Berlind} \& {Weinberg}(2002)}]{Berlind:Weinberg:2002}
{Berlind}, A.~A., \& {Weinberg}, D.~H. 2002, \apj, 575, 587, \dodoi{10.1086/341469}

\bibitem[{{Berti} {et~al.}(2017){Berti}, {Coil}, {Behroozi}, {Eisenstein}, {Bray}, {Cool}, \& {Moustakas}}]{Berti:2017}
{Berti}, A.~M., {Coil}, A.~L., {Behroozi}, P.~S., {et~al.} 2017, \apj, 834, 87, \dodoi{10.3847/1538-4357/834/1/87}

\bibitem[{{Beutler} {et~al.}(2011){Beutler}, {Blake}, {Colless}, {Jones}, {Staveley-Smith}, {Campbell}, {Parker}, {Saunders}, \& {Watson}}]{Beutler:2011}
{Beutler}, F., {Blake}, C., {Colless}, M., {et~al.} 2011, \mnras, 416, 3017, \dodoi{10.1111/j.1365-2966.2011.19250.x}

\bibitem[{{Bianchi} \& {Percival}(2017)}]{Bianchi:Percival:2017}
{Bianchi}, D., \& {Percival}, W.~J. 2017, \mnras, 472, 1106, \dodoi{10.1093/mnras/stx2053}

\bibitem[{Bradbury {et~al.}(2018)Bradbury, Frostig, Hawkins, Johnson, Leary, Maclaurin, Necula, Paszke, Vander{P}las, Wanderman-{M}ilne, \& Zhang}]{JaxGitHub:2018}
Bradbury, J., Frostig, R., Hawkins, P., {et~al.} 2018, {JAX}: composable transformations of {P}ython+{N}um{P}y programs, 0.3.13.
\newblock \url{http://github.com/google/jax}

\bibitem[{{Breiman}(2001)}]{Breiman:2001}
{Breiman}, L. 2001, Machine Learning, 45, 5, \dodoi{10.1023/A:1010933404324}

\bibitem[{{Contreras} {et~al.}(2021){Contreras}, {Chaves-Montero}, {Zennaro}, \& {Angulo}}]{Contreras:2021}
{Contreras}, S., {Chaves-Montero}, J., {Zennaro}, M., \& {Angulo}, R.~E. 2021, \mnras, 507, 3412, \dodoi{10.1093/mnras/stab2367}

\bibitem[{{DESI Collaboration} {et~al.}(2022){DESI Collaboration}, {Abareshi}, {Aguilar}, {Ahlen}, {Alam}, {Alexander}, {Alfarsy}, {Allen}, {Allende Prieto}, {Alves}, \& et~al.}]{DESI:2022}
{DESI Collaboration}, {Abareshi}, B., {Aguilar}, J., {et~al.} 2022, \aj, 164, 207, \dodoi{10.3847/1538-3881/ac882b}

\bibitem[{{DESI Collaboration} {et~al.}(2023){DESI Collaboration}, {Adame}, {Aguilar}, {Ahlen}, {Alam}, {Aldering}, {Alexander}, {Alfarsy}, {Allende Prieto}, {Alvarez}, \& et~al.}]{DESI:2023}
{DESI Collaboration}, {Adame}, A.~G., {Aguilar}, J., {et~al.} 2023, arXiv e-prints, arXiv:2306.06308.
\newblock \doarXiv{2306.06308}

\bibitem[{{Dey} {et~al.}(2019){Dey}, {Schlegel}, {Lang}, {Blum}, {Burleigh}, {Fan}, {Findlay}, {Finkbeiner}, {Herrera}, {Juneau}, {Landriau}, {Levi}, {McGreer}, {Meisner}, {Myers}, {Moustakas}, {Nugent}, {Patej}, {Schlafly}, {Walker}, {Valdes}, {Weaver}, {Y{\`e}che}, {Zou}, {Zhou}, {Abareshi}, {Abbott}, {Abolfathi}, {Aguilera}, {Alam}, {Allen}, {Alvarez}, {Annis}, {Ansarinejad}, {Aubert}, {Beechert}, {Bell}, {BenZvi}, {Beutler}, {Bielby}, {Bolton}, {Brice{\~n}o}, {Buckley-Geer}, {Butler}, {Calamida}, {Carlberg}, {Carter}, {Casas}, {Castander}, {Choi}, {Comparat}, {Cukanovaite}, {Delubac}, {DeVries}, {Dey}, {Dhungana}, {Dickinson}, {Ding}, {Donaldson}, {Duan}, {Duckworth}, {Eftekharzadeh}, {Eisenstein}, {Etourneau}, {Fagrelius}, {Farihi}, {Fitzpatrick}, {Font-Ribera}, {Fulmer}, {G{\"a}nsicke}, {Gaztanaga}, {George}, {Gerdes}, {Gontcho}, {Gorgoni}, {Green}, {Guy}, {Harmer}, {Hernandez}, {Honscheid}, {Huang}, {James}, {Jannuzi}, {Jiang}, {Joyce}, {Karcher}, {Karkar}, {Kehoe}, {Kneib}, {Kueter-Young}, {Lan},
  {Lauer}, {Le Guillou}, {Le Van Suu}, {Lee}, {Lesser}, {Perreault Levasseur}, {Li}, {Mann}, {Marshall}, {Mart{\'\i}nez-V{\'a}zquez}, {Martini}, {du Mas des Bourboux}, {McManus}, {Meier}, {M{\'e}nard}, {Metcalfe}, {Mu{\~n}oz-Guti{\'e}rrez}, {Najita}, {Napier}, {Narayan}, {Newman}, {Nie}, {Nord}, {Norman}, {Olsen}, {Paat}, {Palanque-Delabrouille}, {Peng}, {Poppett}, {Poremba}, {Prakash}, {Rabinowitz}, {Raichoor}, {Rezaie}, {Robertson}, {Roe}, {Ross}, {Ross}, {Rudnick}, {Safonova}, {Saha}, {S{\'a}nchez}, {Savary}, {Schweiker}, {Scott}, {Seo}, {Shan}, {Silva}, {Slepian}, {Soto}, {Sprayberry}, {Staten}, {Stillman}, {Stupak}, {Summers}, {Sien Tie}, {Tirado}, {Vargas-Maga{\~n}a}, {Vivas}, {Wechsler}, {Williams}, {Yang}, {Yang}, {Yapici}, {Zaritsky}, {Zenteno}, {Zhang}, {Zhang}, {Zhou}, \& {Zhou}}]{Dey:2019}
{Dey}, A., {Schlegel}, D.~J., {Lang}, D., {et~al.} 2019, \aj, 157, 168, \dodoi{10.3847/1538-3881/ab089d}

\bibitem[{{Fielder} {et~al.}(2020){Fielder}, {Mao}, {Zentner}, {Newman}, {Wu}, \& {Wechsler}}]{Fielder:2020}
{Fielder}, C.~E., {Mao}, Y.-Y., {Zentner}, A.~R., {et~al.} 2020, \mnras, 499, 2426, \dodoi{10.1093/mnras/staa2851}

\bibitem[{{Foreman-Mackey}(2016)}]{Foreman-Mackey:2016}
{Foreman-Mackey}, D. 2016, The Journal of Open Source Software, 1, 24, \dodoi{10.21105/joss.00024}

\bibitem[{{Foreman-Mackey} {et~al.}(2013){Foreman-Mackey}, {Hogg}, {Lang}, \& {Goodman}}]{Foreman-Mackey:2013}
{Foreman-Mackey}, D., {Hogg}, D.~W., {Lang}, D., \& {Goodman}, J. 2013, \pasp, 125, 306, \dodoi{10.1086/670067}

\bibitem[{{Hearin} {et~al.}(2016){Hearin}, {Zentner}, {van den Bosch}, {Campbell}, \& {Tollerud}}]{Hearin:2016}
{Hearin}, A.~P., {Zentner}, A.~R., {van den Bosch}, F.~C., {Campbell}, D., \& {Tollerud}, E. 2016, \mnras, 460, 2552, \dodoi{10.1093/mnras/stw840}

\bibitem[{{Hubble}(1936)}]{Hubble:1936}
{Hubble}, E.~P. 1936, {Realm of the Nebulae}

\bibitem[{{Hunter}(2007)}]{Hunter:2007}
{Hunter}, J.~D. 2007, Computing in Science and Engineering, 9, 90, \dodoi{10.1109/MCSE.2007.55}

\bibitem[{{Ishiyama} {et~al.}(2021){Ishiyama}, {Prada}, {Klypin}, {Sinha}, {Metcalf}, {Jullo}, {Altieri}, {Cora}, {Croton}, {de la Torre}, {Mill{\'a}n-Calero}, {Oogi}, {Ruedas}, \& {Vega-Mart{\'\i}nez}}]{Ishiyama:2021}
{Ishiyama}, T., {Prada}, F., {Klypin}, A.~A., {et~al.} 2021, \mnras, 506, 4210, \dodoi{10.1093/mnras/stab1755}

\bibitem[{{Kauffmann} {et~al.}(2013){Kauffmann}, {Li}, {Zhang}, \& {Weinmann}}]{Kauffmann:2013}
{Kauffmann}, G., {Li}, C., {Zhang}, W., \& {Weinmann}, S. 2013, \mnras, 430, 1447, \dodoi{10.1093/mnras/stt007}

\bibitem[{{Klypin} {et~al.}(2016){Klypin}, {Yepes}, {Gottl{\"o}ber}, {Prada}, \& {He{\ss}}}]{Klypin:2016}
{Klypin}, A., {Yepes}, G., {Gottl{\"o}ber}, S., {Prada}, F., \& {He{\ss}}, S. 2016, \mnras, 457, 4340, \dodoi{10.1093/mnras/stw248}

\bibitem[{{Landy} \& {Szalay}(1993)}]{Landy:Szalay:1993}
{Landy}, S.~D., \& {Szalay}, A.~S. 1993, \apj, 412, 64, \dodoi{10.1086/172900}

\bibitem[{{Lundberg} \& {Lee}(2017)}]{Lundberg:2017}
{Lundberg}, S., \& {Lee}, S.-I. 2017, arXiv e-prints, arXiv:1705.07874, \dodoi{10.48550/arXiv.1705.07874}

\bibitem[{{Mezini} {et~al.}(2023){Mezini}, {Fielder}, {Zentner}, {Mao}, {Wang}, \& {Wu}}]{Mezini:2023}
{Mezini}, L., {Fielder}, C.~E., {Zentner}, A.~R., {et~al.} 2023, arXiv e-prints, arXiv:2304.13809, \dodoi{10.48550/arXiv.2304.13809}

\bibitem[{{Neal}(2011)}]{Neal:2011}
{Neal}, R. 2011, in Handbook of Markov Chain Monte Carlo, 113--162

\bibitem[{{Pedregosa} {et~al.}(2011){Pedregosa}, {Varoquaux}, {Gramfort}, {Michel}, {Thirion}, {Grisel}, {Blondel}, {M{\"u}ller}, {Nothman}, {Louppe}, {Prettenhofer}, {Weiss}, {Dubourg}, {Vanderplas}, {Passos}, {Cournapeau}, {Brucher}, {Perrot}, \& {Duchesnay}}]{Pedregosa:2012}
{Pedregosa}, F., {Varoquaux}, G., {Gramfort}, A., {et~al.} 2011, Journal of Machine Learning Research, 12, 2825, \dodoi{10.48550/arXiv.1201.0490}

\bibitem[{{Peebles}(1980)}]{Peebles:1980}
{Peebles}, P.~J.~E. 1980, {The large-scale structure of the universe}

\bibitem[{{Penrose}(1955)}]{Penrose:1955}
{Penrose}, R. 1955, Proceedings of the Cambridge Philosophical Society, 51, 406, \dodoi{10.1017/S0305004100030401}

\bibitem[{{Planck Collaboration} {et~al.}(2020){Planck Collaboration}, {Aghanim}, {Akrami}, {Ashdown}, {Aumont}, {Baccigalupi}, {Ballardini}, {Banday}, {Barreiro}, {Bartolo}, {Basak}, {Battye}, {Benabed}, {Bernard}, {Bersanelli}, {Bielewicz}, {Bock}, {Bond}, {Borrill}, {Bouchet}, {Boulanger}, {Bucher}, {Burigana}, {Butler}, {Calabrese}, {Cardoso}, {Carron}, {Challinor}, {Chiang}, {Chluba}, {Colombo}, {Combet}, {Contreras}, {Crill}, {Cuttaia}, {de Bernardis}, {de Zotti}, {Delabrouille}, {Delouis}, {Di Valentino}, {Diego}, {Dor{\'e}}, {Douspis}, {Ducout}, {Dupac}, {Dusini}, {Efstathiou}, {Elsner}, {En{\ss}lin}, {Eriksen}, {Fantaye}, {Farhang}, {Fergusson}, {Fernandez-Cobos}, {Finelli}, {Forastieri}, {Frailis}, {Fraisse}, {Franceschi}, {Frolov}, {Galeotta}, {Galli}, {Ganga}, {G{\'e}nova-Santos}, {Gerbino}, {Ghosh}, {Gonz{\'a}lez-Nuevo}, {G{\'o}rski}, {Gratton}, {Gruppuso}, {Gudmundsson}, {Hamann}, {Handley}, {Hansen}, {Herranz}, {Hildebrandt}, {Hivon}, {Huang}, {Jaffe}, {Jones}, {Karakci}, {Keih{\"a}nen},
  {Keskitalo}, {Kiiveri}, {Kim}, {Kisner}, {Knox}, {Krachmalnicoff}, {Kunz}, {Kurki-Suonio}, {Lagache}, {Lamarre}, {Lasenby}, {Lattanzi}, {Lawrence}, {Le Jeune}, {Lemos}, {Lesgourgues}, {Levrier}, {Lewis}, {Liguori}, {Lilje}, {Lilley}, {Lindholm}, {L{\'o}pez-Caniego}, {Lubin}, {Ma}, {Mac{\'\i}as-P{\'e}rez}, {Maggio}, {Maino}, {Mandolesi}, {Mangilli}, {Marcos-Caballero}, {Maris}, {Martin}, {Martinelli}, {Mart{\'\i}nez-Gonz{\'a}lez}, {Matarrese}, {Mauri}, {McEwen}, {Meinhold}, {Melchiorri}, {Mennella}, {Migliaccio}, {Millea}, {Mitra}, {Miville-Desch{\^e}nes}, {Molinari}, {Montier}, {Morgante}, {Moss}, {Natoli}, {N{\o}rgaard-Nielsen}, {Pagano}, {Paoletti}, {Partridge}, {Patanchon}, {Peiris}, {Perrotta}, {Pettorino}, {Piacentini}, {Polastri}, {Polenta}, {Puget}, {Rachen}, {Reinecke}, {Remazeilles}, {Renzi}, {Rocha}, {Rosset}, {Roudier}, {Rubi{\~n}o-Mart{\'\i}n}, {Ruiz-Granados}, {Salvati}, {Sandri}, {Savelainen}, {Scott}, {Shellard}, {Sirignano}, {Sirri}, {Spencer}, {Sunyaev}, {Suur-Uski}, {Tauber}, {Tavagnacco},
  {Tenti}, {Toffolatti}, {Tomasi}, {Trombetti}, {Valenziano}, {Valiviita}, {Van Tent}, {Vibert}, {Vielva}, {Villa}, {Vittorio}, {Wandelt}, {Wehus}, {White}, {White}, {Zacchei}, \& {Zonca}}]{Planck:2018}
{Planck Collaboration}, {Aghanim}, N., {Akrami}, Y., {et~al.} 2020, \aap, 641, A6, \dodoi{10.1051/0004-6361/201833910}

\bibitem[{{Reddick} {et~al.}(2013){Reddick}, {Wechsler}, {Tinker}, \& {Behroozi}}]{Reddick:2013}
{Reddick}, R.~M., {Wechsler}, R.~H., {Tinker}, J.~L., \& {Behroozi}, P.~S. 2013, \apj, 771, 30, \dodoi{10.1088/0004-637X/771/1/30}

\bibitem[{{Reid} \& {Spergel}(2009)}]{Reid:Spergel:2009}
{Reid}, B.~A., \& {Spergel}, D.~N. 2009, \apj, 698, 143, \dodoi{10.1088/0004-637X/698/1/143}

\bibitem[{{Sato-Polito} {et~al.}(2019){Sato-Polito}, {Montero-Dorta}, {Abramo}, {Prada}, \& {Klypin}}]{Sato-Polito:2019}
{Sato-Polito}, G., {Montero-Dorta}, A.~D., {Abramo}, L.~R., {Prada}, F., \& {Klypin}, A. 2019, \mnras, 487, 1570, \dodoi{10.1093/mnras/stz1338}

\bibitem[{{Sinha} \& {Garrison}(2020)}]{Sinha:2020}
{Sinha}, M., \& {Garrison}, L.~H. 2020, \mnras, 491, 3022, \dodoi{10.1093/mnras/stz3157}

\bibitem[{{Smith} {et~al.}(2019){Smith}, {He}, {Cole}, {Stothert}, {Norberg}, {Baugh}, {Bianchi}, {Wilson}, {Brooks}, {Forero-Romero}, {Moustakas}, {Percival}, {Tarle}, \& {Wechsler}}]{Smith:2019}
{Smith}, A., {He}, J.-h., {Cole}, S., {et~al.} 2019, \mnras, 484, 1285, \dodoi{10.1093/mnras/stz059}

\bibitem[{{Storey-Fisher} {et~al.}(2022){Storey-Fisher}, {Tinker}, {Zhai}, {DeRose}, {Wechsler}, \& {Banerjee}}]{Storey-Fisher:2022}
{Storey-Fisher}, K., {Tinker}, J., {Zhai}, Z., {et~al.} 2022, arXiv e-prints, arXiv:2210.03203, \dodoi{10.48550/arXiv.2210.03203}

\bibitem[{{Vakili} \& {Hahn}(2019)}]{Vakili:Hahn:2019}
{Vakili}, M., \& {Hahn}, C. 2019, \apj, 872, 115, \dodoi{10.3847/1538-4357/aaf1a1}

\bibitem[{{van der Walt} {et~al.}(2011){van der Walt}, {Colbert}, \& {Varoquaux}}]{vanderWalt:2011}
{van der Walt}, S., {Colbert}, S.~C., \& {Varoquaux}, G. 2011, Computing in Science and Engineering, 13, 22, \dodoi{10.1109/MCSE.2011.37}

\bibitem[{{Virtanen} {et~al.}(2020){Virtanen}, {Gommers}, {Oliphant}, {Haberland}, {Reddy}, {Cournapeau}, {Burovski}, {Peterson}, {Weckesser}, {Bright}, {van der Walt}, {Brett}, {Wilson}, {Millman}, {Mayorov}, {Nelson}, {Jones}, {Kern}, {Larson}, {Carey}, {Polat}, {Feng}, {Moore}, {Vand erPlas}, {Laxalde}, {Perktold}, {Cimrman}, {Henriksen}, {Quintero}, {Harris}, {Archibald}, {Ribeiro}, {Pedregosa}, {van Mulbregt}, \& {SciPy 1. 0 Contributors}}]{Virtanen:2020}
{Virtanen}, P., {Gommers}, R., {Oliphant}, T.~E., {et~al.} 2020, Nature Methods, 17, 261, \dodoi{10.1038/s41592-019-0686-2}

\bibitem[{{Wang} {et~al.}(2022){Wang}, {Mao}, {Zentner}, {Guo}, {Lange}, {van den Bosch}, \& {Mezini}}]{Wang:2022}
{Wang}, K., {Mao}, Y.-Y., {Zentner}, A.~R., {et~al.} 2022, \mnras, 516, 4003, \dodoi{10.1093/mnras/stac2465}

\bibitem[{{Wang} {et~al.}(2019){Wang}, {Mao}, {Zentner}, {van den Bosch}, {Lange}, {Schafer}, {Villarreal}, {Hearin}, \& {Campbell}}]{Wang:2019}
---. 2019, \mnras, 488, 3541, \dodoi{10.1093/mnras/stz1733}

\bibitem[{{White}(1979)}]{White:1979}
{White}, S.~D.~M. 1979, \mnras, 186, 145, \dodoi{10.1093/mnras/186.2.145}

\bibitem[{{Yuan} {et~al.}(2021){Yuan}, {Hadzhiyska}, {Bose}, {Eisenstein}, \& {Guo}}]{Yuan:2021}
{Yuan}, S., {Hadzhiyska}, B., {Bose}, S., {Eisenstein}, D.~J., \& {Guo}, H. 2021, \mnras, 502, 3582, \dodoi{10.1093/mnras/stab235}

\bibitem[{{Zehavi} {et~al.}(2005){Zehavi}, {Zheng}, {Weinberg}, {Frieman}, {Berlind}, {Blanton}, {Scoccimarro}, {Sheth}, {Strauss}, {Kayo}, {Suto}, {Fukugita}, {Nakamura}, {Bahcall}, {Brinkmann}, {Gunn}, {Hennessy}, {Ivezi{\'c}}, {Knapp}, {Loveday}, {Meiksin}, {Schlegel}, {Schneider}, {Szapudi}, {Tegmark}, {Vogeley}, {York}, \& {SDSS Collaboration}}]{Zehavi:2005}
{Zehavi}, I., {Zheng}, Z., {Weinberg}, D.~H., {et~al.} 2005, \apj, 630, 1, \dodoi{10.1086/431891}

\bibitem[{{Zentner} {et~al.}(2019){Zentner}, {Hearin}, {van den Bosch}, {Lange}, \& {Villarreal}}]{Zentner:2019}
{Zentner}, A.~R., {Hearin}, A., {van den Bosch}, F.~C., {Lange}, J.~U., \& {Villarreal}, A. 2019, \mnras, 485, 1196, \dodoi{10.1093/mnras/stz470}

\bibitem[{{Zentner} {et~al.}(2005){Zentner}, {Kravtsov}, {Gnedin}, \& {Klypin}}]{Zentner:2005}
{Zentner}, A.~R., {Kravtsov}, A.~V., {Gnedin}, O.~Y., \& {Klypin}, A.~A. 2005, \apj, 629, 219, \dodoi{10.1086/431355}

\bibitem[{{Zheng} {et~al.}(2007){Zheng}, {Coil}, \& {Zehavi}}]{Zheng:2007}
{Zheng}, Z., {Coil}, A.~L., \& {Zehavi}, I. 2007, \apj, 667, 760, \dodoi{10.1086/521074}

\bibitem[{{Zwicky}(1957)}]{Zwicky:1957}
{Zwicky}, F. 1957, {Morphological astronomy}

\end{thebibliography}

\end{document}